\documentclass[journal]{IEEEtran}

\usepackage[font={footnotesize}]{caption}
\usepackage{subcaption}
\usepackage{cite}
\usepackage{balance}
\usepackage{amsmath,amssymb,amsfonts}
\usepackage{algorithmic}
\usepackage{graphicx}
\usepackage{textcomp}
\usepackage{xcolor}
\usepackage{tikz}
\usepackage{pgfplots} 
\usepackage{tikzscale}
\usepackage{pgfplots}
\usepackage{cite}
\usepackage{amsmath,amssymb,amsfonts}
\usepackage{algorithmic}
\usepackage{textcomp}
\usepackage{accents} 
\usepackage{color}
\usepackage{comment}

\usepackage{acro}

\DeclareAcronym{5g}{
short=5G,
long= fifth generation,
}
\DeclareAcronym{6g}{
short=6G,
long= sixth generation,
}
\DeclareAcronym{3d}{
short=3D,
long= three-dimensional,
}
\DeclareAcronym{em}{
short=EM,
long= electromagnetic,
}
\DeclareAcronym{aod}{
short=AOD,
long= angle-of-departure,
}
\DeclareAcronym{aosa}{
short=AOSA,
long= array-of-subarray,
}
\DeclareAcronym{adod}{
short=ADOD,
long= angle-difference-of-departure,
}
\DeclareAcronym{aoa}{
short=AOA,
long= angle-of-arrival,
}
\DeclareAcronym{adc}{
short=ADC,
long= analog to digital converter,
}
\DeclareAcronym{aeb}{
short=AEB,
long= angle error bound,
}
\DeclareAcronym{av}{
short=AV,
long= autonomous vehicle,
}
\DeclareAcronym{bs}{
short=BS,
long= base station,
}
\DeclareAcronym{bse}{
short=BSE,
long= beam split effect,
}
\DeclareAcronym{csi}{
short=CSI,
long= channel state information,
}
\DeclareAcronym{cfo}{
short=CFO,
long= carrier frequency offset,
}
\DeclareAcronym{ceb}{
short=CEB,
long= clock error bound,
}
\DeclareAcronym{coa}{
short=COA,
long= curvature-of-arrival,
}
\DeclareAcronym{crb}{
short=CRB,
long= Cram\'er-Rao lower bound,
}
\DeclareAcronym{ccrb}{
short=CCRB,
long= constrained Cram\'er-Rao bound,
}
\DeclareAcronym{SP}{
short=SP,
long= scatter point,
}
\DeclareAcronym{cmos}{
short=CMOS,
long= complementary metal-oxide-semiconductor,
}
\DeclareAcronym{crlb}{
short=CRLB,
long= Cram\'er-Rao lower bound,
}
\DeclareAcronym{cdf}{
short=CDF,
long= cumulative distribution function,
}
\DeclareAcronym{cp}{
short=CP,
long= cyclic prefix,
}
\DeclareAcronym{dac}{
short=DAC,
long= digital to analog converter,
}
\DeclareAcronym{dfl}{
short=DFL,
long= device-free localization,
}
\DeclareAcronym{dmimo}{
short=D-MIMO,
long= distributed MIMO,
}
\DeclareAcronym{dlprs}{
short=DL-PRS,
long= downlink positioning reference signal,
}
\DeclareAcronym{d2d}{
short=D2D,
long= device-to-device,
}

\DeclareAcronym{dftsofdm}{
short=DFT-s-OFDM,
long= discrete-Fourier-transform spread OFDM,
}
\DeclareAcronym{dft}{
short=DFT,
long= discrete-Fourier-transform,
}
\DeclareAcronym{dl}{
short=DL,
long= deep learning,
}
\DeclareAcronym{gps}{
short=GPS,
long= global positioning system,
}
\DeclareAcronym{hwi}{
short=HWI,
long= hardware impairment,
}
\DeclareAcronym{hemt}{
short=HEMT,
long= high electron mobility transistor,
}
\DeclareAcronym{hbt}{
short=HBT,
long= heterojunction bipolar transistors,
}
\DeclareAcronym{iot}{
short=IoT,
long= internet of things,
}
\DeclareAcronym{isac}{
short=ISAC,
long= integrated sensing and communication,
}
\DeclareAcronym{iqi}{
short=IQI,
long= in-phase and quadrature imbalance,
}
\DeclareAcronym{ia}{
short=IA,
long= initial access,
}
\DeclareAcronym{kpi}{
short=KPI,
long= key performance indicator,
}
\DeclareAcronym{kf}{
short=KF,
long= Kalman filter,
}
\DeclareAcronym{ekf}{
short=EKF,
long= extended Kalman filter,
}
\DeclareAcronym{ukf}{
short=UKF,
long= unscented Kalman filter,
}
\DeclareAcronym{ckf}{
short=CKF,
long= cubature Kalman filter,
}
\DeclareAcronym{fim}{
short=FIM,
long= Fisher information matrix,
}
\DeclareAcronym{pf}{
short=PF,
long= particle filter,
}
\DeclareAcronym{lb}{
short=LB,
long= lower bound,
}
\DeclareAcronym{lse}{
short=LSE,
long= least-square estimator,
}
\DeclareAcronym{lo}{
short=LO,
long= local oscillator,
}
\DeclareAcronym{mc}{
short=MC,
long= mutual coupling,
}
\DeclareAcronym{mac}{
short=MAC,
long= medium access control,
}
\DeclareAcronym{meb}{
short=MEB,
long= mapping error bound,
}
\DeclareAcronym{ml}{
short=ML,
long= machine learning,
}
\DeclareAcronym{mcrb}{
short=MCRB,
long= misspecified Cram\'er-Rao bound,
}
\DeclareAcronym{mds}{
short=MDS,
long= multidimensional scaling ,
}
\DeclareAcronym{mimo}{
short=MIMO,
long= multiple-input-multiple-output,
}
\DeclareAcronym{mm}{
short=MM,
long= mismatched model,
}
\DeclareAcronym{mpc}{
short=MPC,
long= multipath components,
}
\DeclareAcronym{mmwave}{
short=mmWave,
long= millimeter wave,
}
\DeclareAcronym{mmle}{
short=MMLE,
long= mismatched maximum likelihood estimation,
}
\DeclareAcronym{mems}{
short=MEMS,
long= micro-electro-mechanical system,
}
\DeclareAcronym{mle}{
short=MLE,
long= maximum likelihood estimation,
}
\DeclareAcronym{nlos}{
short=NLOS,
long= non-line-of-sight,
}
\DeclareAcronym{ofdm}{
short=OFDM,
long= orthogonal frequency-division multiplexing,
}
\DeclareAcronym{oeb}{
short=OEB,
long= orientation error bound,
}
\DeclareAcronym{otfs}{
short=OTFS,
long= orthogonal time-frequency space,
}
\DeclareAcronym{pdf}{
short=PDF,
long= probability density function,
}
\DeclareAcronym{papr}{
short=PAPR,
long= peak-to-average-power ratio,
}
\DeclareAcronym{pan}{
short=PAN,
long= power amplifier nonlinearity,
}
\DeclareAcronym{pa}{
short=PA,
long= power amplifier,
}
\DeclareAcronym{ps}{
short=PS,
long= phase shifter,
}
\DeclareAcronym{pn}{
short=PN,
long= phase noise,
}
\DeclareAcronym{poa}{
short=POA,
long= phase-of-arrival,
}
\DeclareAcronym{pwm}{
short=PWM,
long= planar wave model,
}
\DeclareAcronym{pdoa}{
short=PDOA,
long= phase-difference-of-arrival,
}
\DeclareAcronym{prs}{
short=PRS,
long= positioning reference signals,
}
\DeclareAcronym{peb}{
short=PEB,
long= position error bound,
}
\DeclareAcronym{rnn}{
short=RNN,
long= recurrent neural network,
}
\DeclareAcronym{rl}{
short=RL,
long= reinforcement learning,
}
\DeclareAcronym{rmse}{
short=RMSE,
long= root-mean-square- error,
}
\DeclareAcronym{rfid}{
short=RFID,
long= radio frequency identification,
}
\DeclareAcronym{ris}{
short=RIS,
long= reconfigurable intelligent surface,
}
\DeclareAcronym{hris}{
short=HRIS,
long= hybrid reconfigurable intelligent surface,
}
\DeclareAcronym{rss}{
short=RSS,
long= received signal strength,
}
\DeclareAcronym{rtt}{
short=RTT,
long= round-trip time,
}
\DeclareAcronym{sm}{
short=SM,
long= standard model,
}
\DeclareAcronym{sige}{
short=SiGe,
long= silicon-germanium,
}
\DeclareAcronym{spp}{
short=SPP,
long= surface plasmon polariton,
}
\DeclareAcronym{rf}{
short=RF,
long= radio frequency,
}
\DeclareAcronym{rfc}{
short=RFC,
long= radio frequency chain,
}
\DeclareAcronym{ff}{
short=FF,
long= far field,
}
\DeclareAcronym{los}{
short=LOS,
long= line of sight,
}
\DeclareAcronym{nn}{
short=NF,
long= near field,
}
\DeclareAcronym{sa}{
short=SA,
long= subarray,
}
\DeclareAcronym{sota}{
short=SOTA,
long= state-of-the-art,
}
\DeclareAcronym{swm}{
short=SWM,
long= spherical wave model,
}
\DeclareAcronym{slam}{
short=SLAM,
long= simultaneous localization and mapping,
}
\DeclareAcronym{tm}{
short=TM,
long= true model,
}
\DeclareAcronym{toa}{
short=TOA,
long= time-of-arrival,
}
\DeclareAcronym{tof}{
short=TOF,
long= time-of-flight,
}
\DeclareAcronym{tdoa}{
short=TDOA,
long= time-difference-of-arrival,
}
\DeclareAcronym{thz}{
short=THz,
long= terahertz,
}
\DeclareAcronym{ue}{
short=UE,
long= user equipment,
}
\DeclareAcronym{ummimo}{
short=UM-MIMO,
long= ultra-massive multi-input-multi-output,
}
\DeclareAcronym{vlp}{
short=VLP,
long= visible light positioning,
}
\DeclareAcronym{veb}{
short=VEB,
long= velocity error bound,
}
\DeclareAcronym{vlc}{
short=VLC,
long= visible light communication,
}
\DeclareAcronym{ula}{
short=ULA,
long= uniform linear array,
}
\DeclareAcronym{upa}{
short=UPA,
long= uniform planar array,
}
\DeclareAcronym{wlan}{
short=WLAN,
long= wireless local area network,
}

\usepackage{pgfplots}
\usepackage{tikz}
\usetikzlibrary{calc}
\makeatletter
\newcommand{\gettikzxy}[3]{%
  \tikz@scan@one@point\pgfutil@firstofone#1\relax
  \edef#2{\the\pgf@x}%
  \edef#3{\the\pgf@y}%
}

\def\BibTeX{{\rm B\kern-.05em{\sc i\kern-.025em b}\kern-.08em
    T\kern-.1667em\lower.7ex\hbox{E}\kern-.125emX}}
    
\begin{document}

\title{
Joint 3D User and 6D Hybrid Reconfigurable Intelligent Surface Localization}

\author{Reza~Ghazalian,~\IEEEmembership{Member,~IEEE,} 
George~C.~Alexandropoulos,~\IEEEmembership{Senior Member,~IEEE,}
Gonzalo~Seco-Granados,~\IEEEmembership{Fellow,~IEEE,}
Henk~Wymeersch,~\IEEEmembership{Senior Member,~IEEE,} and~Riku~J{\"a}ntti,~\IEEEmembership{Senior~Member,~IEEE}
\thanks{This work has been funded in part by the Academy of Finland Profi-5 under the grant number 326346, ULTRA under grant number 328215, the EU H2020 RISE-6G project under grant number 10101701, the SNS JU TERRAMETA project under the EU's Horizon Europe research and innovation programme under grant number 101097101, including top-up funding by UKRI under the UK government's Horizon Europe funding guarantee, the ICREA Academia Program, and the Spanish R+D project PID2020-118984GB-I00.}
\thanks{R. Ghazalian and R. Jäntti are with the Department of Information and Communications Engineering, School of Electrical Engineering
of Electrical Engineering, Aalto University, 02150 Espoo, Finland (\{reza.ghazalian, riku.jantti\}@aalto.fi).}
\thanks{H. Wymeersch is with the Department of Electrical Engineering, Chalmers University of Technology, 412 58 Gothenburg, Sweden (emails: henkw@chalmers.se).}
\thanks{G. C. Alexandropoulos is with the Department of Informatics and Telecommunications, National and Kapodistrian University of Athens, 15784 Athens, Greece (e-mail: alexandg@di.uoa.gr).}
\thanks{G. Seco-Granados is with the Department of Telecommunications and
Systems Engineering, Universitat Autònoma de Barcelona, 08193 Barcelona,
Spain (e-mail: gonzalo.seco@uab.cat).}
\vspace{-0.5cm}}

\maketitle
\begin{abstract}
The latest assessments of the emerging technologies for \acp{ris} have indicated the concept's significant potential for localization and sensing, either as individual or simultaneously realized tasks. However, in the vast majority of those studies, the RIS state (i.e., its position and rotation angles) is required to be known a priori.
In this paper, we address the problem of the joint three-dimensional (3D) localization of a hybrid RIS (HRIS) and a user. The most cost- and power-efficient hybrid version of an RIS is equipped with a single reception radio-frequency chain and meta-atoms capable of simultaneous reconfigurable reflection and sensing. This dual functionality is controlled by adjustable power splitters embedded at each hybrid meta-atom. Focusing on a downlink scenario where a multi-antenna base station transmits multi-carrier signals to a user via an HRIS, we propose a multi-stage approach to jointly estimate the metasurface's 3D position and 3D rotation matrix (i.e., 6D parameter estimation) as well as the user's 3D position. Our simulation results verify the validity of the proposed estimator via extensive comparisons of the root-mean-square error of the state estimations with the \acf{crb}, which is analytically derived. Furthermore, it is showcased that there exists an optimal \ac{hris} power splitting ratio for the desired multi-parameter estimation problem. We also study the robustness of the proposed method in the presence of scattering points in the wireless propagation environment. 
\end{abstract}


\section{Introduction}
\IEEEPARstart{R}{Econfigurable} intelligent surfaces (\acp{ris}) are recently being extensively studied as an enabling technology for the upcoming \acf{6g} of wireless systems~\cite{huang2019reconfigurable,Marco2019}. An \ac{ris} is a planar surface made of sub-wavelength unit cells with controllable \acf{em} properties~\cite{wu2019towards,alexandg_2021}. In essence, \acp{ris} can modify wave characteristics such as phase, amplitude, frequency, and even polarization, offering radio propagation control~\cite{WavePropTCCN,RIS_space_shift_keying,RIS_reflection_modulation}. Smart wireless environments can be achieved via this characteristic of \acp{ris}, which provides coverage extension and localization, as well as sensing capabilities~\cite{wymeersch2020radio,RIS_ISAC_SPM}. Hence, \acp{ris} are expected to be the vital enabler for the \ac{6g} of wireless systems~\cite{RISE6G_COMMAG,Alexandropoulos2022Pervasive}, where joint communications and localization is expected to aid various use cases~\cite{saad2019vision}, such as connected robotics, autonomous systems, and other immersive applications~\cite{RIS_metaverse}. 
There have been several studies on \acp{ris}'s utility in radio localization by developing localization algorithms or deriving \acf{crb}, see, e.g.,~\cite{abu2021near,keykhosravi2021siso,keykhosravi2022ris,alexandropoulos2022localization,zhang2022hybrid,gaudreauand2022localization,gan2022near,rinchi2022compressive,zhang2021metalocalization,mylonopoulos2022active,dardari2021nlos,elzanaty2021reconfigurable,ghazalian2022bi,ghazalian2023joint,keykhosravi2021semi,Multiple_passive_RIS}. 
In those works, \acp{ris} have been successfully deployed for \acf{ue} radio localization in two different general scenarios. In the first scenario, except for the \ac{bs}, the \ac{ris} acts as anchor with known state, i.e., location and orientation. These kind of systems are known as \textit{RIS-aided} or \textit{RIS-enabled localization} \cite{keykhosravi2022leveraging}, and exploit reflected signals from \acp{ris} to improve or assist \ac{ue} localization. In the second scenario, a \ac{ue} can use an RIS(s) to enable its  localization, where the RIS(s)'s state is unknown and needs to be estimated. This problem is known as \textit{RIS localization}, which one can see as a
\textit{bi-static sensing} or a calibration problem~\cite{ghazalian2022bi}.

RIS-aided \ac{ue} localization has been intensely studied in recent years, ranging from two-dimensional (2D) to 3D scenarios under either \acf{ff} or \acf{nn} operating conditions, as well as indoor and outdoor scenarios~\cite{huang2019spawc,Samarakoon_2020,abu2021near,keykhosravi2021siso,keykhosravi2022ris,alexandropoulos2022localization,zhang2022hybrid,gaudreauand2022localization,gan2022near,rinchi2022compressive,zhang2021metalocalization,mylonopoulos2022active,dardari2021nlos,elzanaty2021reconfigurable,STAR_RIS_loc,FD_HMIMO_2023,RIS_partially,FD_nfRIS_2023,Nlos_DMA}.
\ac{ue} localization under the \ac{nn} condition via analyzing the \ac{crb} has been studied in~\cite{abu2021near} for passive \ac{ris}, and for hybrid \ac{ris} (\ac{hris}) with \ac{los} blockage in~\cite{zhang2022hybrid}. Moreover, the possibility of \ac{ue} positioning under \acf{nlos} harsh propagation conditions has been shown~\cite{Samarakoon_2020,dardari2021nlos}, in which the \ac{ris} is equipped with a large number of reflecting elements. In~\cite{gan2022near}, the authors showed that the \acf{fim} on the \ac{ue} position estimation, in the case it operates in the ~\ac{nn} regime, quadratically increases with the size of the \ac{ris}. Furthermore, a \ac{ue} localization algorithm based on compressed sensing techniques for an uplink \ac{nn} scenario was presented in~\cite{rinchi2022compressive}. 3D \ac{ue} localization and synchronization in the \ac{ff} scenario has been studied in~\cite{keykhosravi2021siso,keykhosravi2022ris}, where \ac{ue} mobility and the spatial-wideband effect were taken into account. In~\cite{gaudreauand2022localization,Multiple_passive_RIS}, multiple \acp{ris} were adopted for \ac{ue} localization. In~\cite{gaudreauand2022localization}, the \acp{ris} modulate an impinging unmodulated carrier and time-difference-of-arrival is calculated at the receivers to estimate the UE location. The localization approach in~\cite{Multiple_passive_RIS} considered narroband transmission and was based in \ac{aoa} estimation. In~\cite{elzanaty2021reconfigurable}, the potential of using an \ac{ris} for \ac{ue} orientation estimation, besides its location estimation, was studied via an \ac{crb} analysis. 

The problem of RIS localization, which assumes that the \ac{ris} state (i.e., position and/or orientation) is unknown, has been rarely discussed.  While lack of knowledge of the RIS position and orientation generally does not affect RIS-aided communication, it prevents RIS-aided UE localization \cite{liu2023integrated}. The RIS position or orientation may be unknown for a number of reasons, including poor calibration, lack of knowledge of the environment map in which the RIS is placed, or due to mobility of the RIS itself~\cite{chen2022efficient,RIS_UAV_2023}. 
The relevant works are only~\cite{ghazalian2022bi,keykhosravi2021semi} that focused on a passive \ac{ris}. In these works, the locations of the transmitters and receivers are assumed to be known, which requires some overhead for infrastructure calibration. The estimation of \ac{aoa} and \ac{aod} separately at the \ac{ris} offers a promising way for \ac{ris} localization jointly with the localization of one of the transmitter and receiver~\cite{ghazalian2023joint}, reducing the aforementioned calibration overhead. However, one cannot separately estimate the \ac{aoa} and \ac{aod} at the passive \ac{ris}. Adding sensing capability to the \ac{ris} can be a promising solution to overcome this challenge~\cite{ghazalian2023joint}, which was first proposed in~\cite{alexandropoulos2020hardware} for individual channel estimation. This version of an \ac{ris} was then extended in~\cite{alexandropoulos2021hybrid,alamzadeh2021reconfigurable,zhang2022channel} to a HRIS that is capable of simultaneous reflection and sensing. According to this RIS hardware architecture, waveguides feed the incident signals at each hybrid meta-atom to reception (RX) \ac{rf} chains, which is connected to a baseband unit. This reception mechanism effectively acts as an analog combiner whose phase shifts can be dynamically optimized~\cite{alexandropoulos2021hybrid}. Very recently, \cite{zhang2022hybrid} studied UE localization in an NF scenario, considering multiple RX \ac{rf} chains at the HRIS and that its state is precisely known.  HRIS-assisted MIMO radar-communication systems has been investigated, where the HRIS simultaneously reflected communications signals and received radar echos~\cite{liu2023hybrid}.
However, HRISs with a single RX \ac{rf} chain in localization problems, where their states are unknown, have not yer been investigated.

In this paper, we extend our recent work in~\cite{ghazalian2023joint}, which addressed joint 2D \ac{ue} and \ac{hris} localization, to joint 3D \acp{ue} and \ac{hris} localization under \ac{ff} conditions, including the 3D rotation matrix estimation for the HRIS. The main contributions of this paper are summarized as follows.
\begin{itemize}
\item We devise a multi-stage algorithm for joint multiple-input-single-output (MISO) 3D localization
and synchronization of a \ac{ue} and an \ac{hris} under \ac{ff} scenarios. In the first step, we estimate the geometric parameters of the associated channels, i.e., delay, gain, \ac{aoa}/\ac{aod} at \ac{hris}, and \ac{aod} at the BS. The \ac{aoa}/\ac{aod} estimation at the HRIS is enabled by its single RX RF chain. Using these estimations at a further stage, we calculate the 3D position and clock bias of the \ac{ue} and the \ac{hris}. At the final step of the proposed approach, we make use of the positions' estimates, the \ac{aoa} at the HRIS from the BS and the \ac{aod} from the HRIS to the \ac{ue} to estimate the 3D rotation matrix of the HRIS.  
  \item We investigate the role of several system parameters (the transmit power at the BS, the power splitting ratio at the \ac{hris}, and the presence of scatterers) on the error of the  proposed estimation approach via extensive simulations, and through comparisons with the estimator's \ac{crb}, which is derived for this purpose. In this investigation, we consider \ac{hris} phase profiles from the \ac{dft}. Our simulation results showcase the efficiency of the proposed estimatin algorithm, which attains the corresponding \ac{crb} at high transmit power levers. The critical role of the \ac{hris} power splitting ratio on the localization performance is also unveiled. We finally show that the proposed approach is robust to the presence of the scattering points (SPs). 
\end{itemize}

The rest of this paper is organized as follows: In Section~\ref{sec_sys_sig_model},
the system and signal models are described, while the Fisher information of the estimations is detailed in Section~\ref{Sec.FIM}. The proposed estimation algorithm is presented in Section~\ref{sec_estimator}. In Section~\ref{Sec_simulation}, we numerically evaluate the \acf{rmse} of the parameters' estimation and compare it with the respective \acp{crb}. Section~\ref{Sec_conclusion} concludes the paper.
\subsubsection*{Notation}
Vectors and matrices are indicated by lowercase and uppercase bold letters, respectively. The element in the $i$-th row and
$j$-th column of matrix $\mathbf{A}$ is denoted by $\left[\mathbf{A}\right]_{i,j}$ and $\det(\mathbf{A})$ returns $\mathbf{A}$'s determinant. The sub-index $i:j$ determines all the elements between $i$ and $j$. The complex conjugate, Hermitian, transpose, and Moore–Penrose inverse operators are represented by $\left( .\right)^*$, $\left( .\right)^{\mathsf{H}}$, $\left( .\right)^\top$, and $\left( .\right)^\dag$, respectively. $\Vert.\Vert$ calculates the norm of vectors or Frobenius norm of matrices. $\odot$ and $\otimes$ indicate the element-wise and Kronecker products, respectively. $\mathbb{R}$ and $\mathbb{C}$ are the real and complex number sets, $\Re\{\cdot\}$ and $\Im\{\cdot\}$ give the real and imaginary parts of a complex number, and $\jmath=\sqrt{-1}$. $\mathbf{1}_K$ and $\mathbf{0}_K$ are column vectors  with length $K$ comprising all ones and zeros, respectively. The functions $\text{atan2}(y,x)$ and $\text{acos(x)}$ are the four-quadrant inverse tangent and inverse cosine functions, respectively. 
\begin{figure}[t]
\centering
\includegraphics[width=0.8\linewidth]{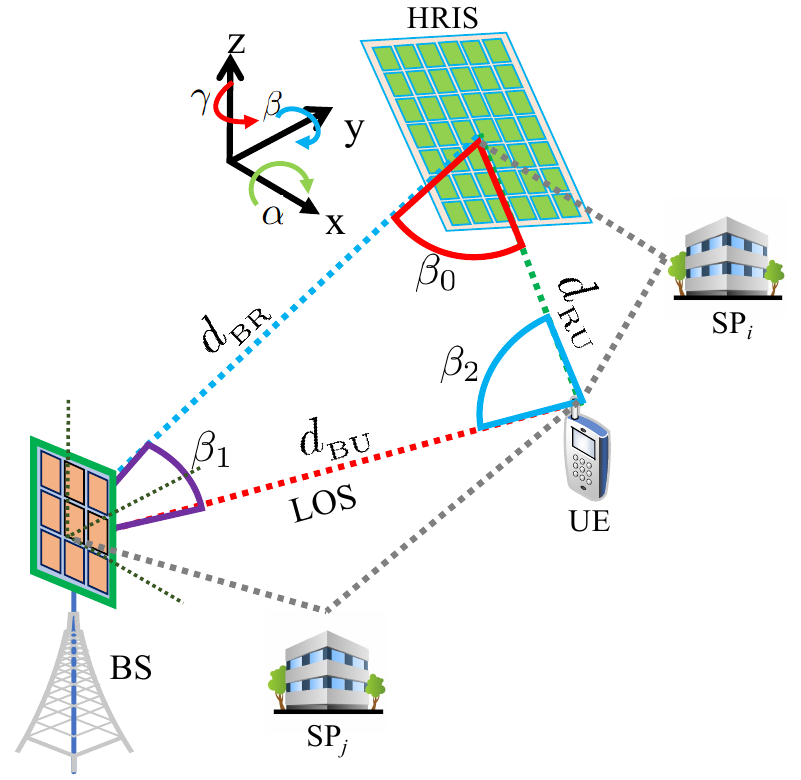}
\caption{The considered wireless system comprising a multi-antenna BS, a single-antenna UE, and a single-RX-RF HRIS. The 3D location of the UE and the 6D state of the HRIS are unknown.}
\label{fig:Scenario}
\end{figure}
\vspace{5mm}
\section{System and Signal Models}\label{sec_sys_sig_model}
In this section, we introduce the considered HRIS-empowered system model as well as the models for the received signals and the wireless channel that will be deployed for the proposed localization approach.

\subsection{System Setup}
Consider the wireless system scenario in Fig.~\ref{fig:Scenario} consisting of one $M_{_\text{B}}$-antenna BS with a known location $\mathbf{p}_\text{B}\in \mathbb{R}^3$, one single-antenna UE with an unknown location $\mathbf{p}_\text{U}\in \mathbb{R}^3$, and an HRIS with unknown location $\mathbf{p}_\text{R}\in \mathbb{R}^3$ and unknown orientation\footnote{One can extend the scenario to the multi-user case, which would improve the localization performance. As the positioning method is working in the DL, one can locate any other user, and this would actually improve the overall performance because the measurement from all users would contribute to the determination of the HRIS state.
}. We consider the downlink scenario where the BS sends $T$ orthogonal frequency division multiplexing (OFDM) symbols over time via $K$ subcarriers. We assume that all the associated channels remain constant during each transmission time slot. To model the HRIS orientation, we use a rotation matrix $3\times3$ matrix $\mathbf{R}$ that determines local coordinate frames, which belongs in the special orthogonal group SO(3), i.e., orthogonal matrices with unit-valued determinant. 
In particular,
we define a reference orientation where the axes are in the same direction as those of the global coordinate frame, as shown in Fig.~\ref{fig:Scenario}. We assume that the global coordinate system is aligned with the BS local coordinate system. Accordingly, we can express the HRIS rotation matrix as follows:
\begin{align}\label{eq:rot-matrix}
 \mathbf{R}= \mathbf{R}_{\alpha}(z)\mathbf{R}_{\beta}(y)\mathbf{R}_{\gamma}(x),
\end{align}
where $\mathbf{R}_{\beta}(z)$, $\mathbf{R}_{\beta}(y)$, and $\mathbf{R}_{\beta}(x)$ represent the rotation matrices w.r.t. $z$-axis, $y$-axis, and $x$-axis, respectively, and are expressed as follows:
\begin{subequations}\label{eq:rotation_zyx}
\begin{equation}\label{eq:rot-matriz}
 \mathbf{R}_{\alpha}(z) = \begin{bmatrix} \cos{\alpha}&\sin{\alpha}& 0\\-\sin{\alpha}&\cos{\alpha}&0 \\ 0 & 0 & 1  \end{bmatrix},
 \end{equation}
 \begin{equation}\label{eq:rot-matriy}
 \mathbf{R}_{\beta}(y) = \begin{bmatrix} \cos{\beta}&0&\sin{\beta}\\0&1&0\\-\sin{\beta}&0&\cos{\beta}\end{bmatrix},
 \end{equation}
 \begin{equation}\label{eq:rot-matrixx}
 \mathbf{R}_{\gamma}(x) = \begin{bmatrix} 1&0&0\\ 0&\cos{\gamma}&-\sin{\gamma}\\0&\sin{\gamma}&\cos{\gamma}\end{bmatrix}.
 \end{equation}
\end{subequations}
The direction vectors from the HRIS to the BS and the UE in the local coordinate system can be respectively obtained as:
\begin{subequations}\label{eq:direction_HRIS}
\begin{equation}\label{eq:direction_HRIS-BS}
\mathbf{q}_\text{RB}= \mathbf{R}^\top (\mathbf{p}_\text{B}-\mathbf{p}_\text{R})/\Vert\mathbf{p}_\text{B}-\mathbf{p}_\text{R}\Vert,
\end{equation}
\begin{equation}\label{eq:direction_HRIS_UE}
\mathbf{q}_\text{RU}= \mathbf{R}^\top (\mathbf{p}_\text{U}-\mathbf{p}_\text{R})/\Vert\mathbf{p}_\text{U}-\mathbf{p}_\text{R}\Vert.
 \end{equation}
\end{subequations}
Similarly, we respectively write the direction vectors from the BS to the HRIS and the UE in the BS coordinate system as:
\begin{subequations}\label{eq:direction_BS}
 \begin{equation}\label{eq:direction_BS-RIS}
 \mathbf{q}_\text{BR}= (\mathbf{p}_\text{R}-\mathbf{p}_\text{B})/\Vert\mathbf{p}_\text{R}-\mathbf{p}_\text{B}\Vert,
 \end{equation}
 \begin{equation}\label{eq:direction_BS-UE}
 \mathbf{q}_\text{BU}=  (\mathbf{p}_\text{U}-\mathbf{p}_\text{B})/\Vert\mathbf{p}_\text{U}-\mathbf{p}_\text{B}\Vert.
 \end{equation}
\end{subequations}
Moreover, the HRIS and UE are assumed unsynchronized with the BS, leading to the unknown clock biases $b_\text{R}$ and $b_\text{U}$ at the HRIS and UE, respectively, with respect to the BS. Therefore, in addition to the HRIS location and UE position, the latter clock bias components need to be estimated.

The BS is assumed equipped with a uniform planar array (UPA) with $M_{_\text{B}} = M_{_\text{B}}^r \times M_{_\text{B}}^c$ elements.\footnote{Using a single-antenna BS or a BS with a uniform linear array will make the targeted estimation problem infeasible, due to the fact that there will not be enough measurements for this task. However, when deploying a UPA at the BS, four additional measurements are feasible: a 2D AOD measurement at the BS from the UE and the HRIS. For instance, using a single-antenna BS, we will have seven measurements, i.e., 2D AOA and 2D AOD at the RIS from/to the BS/UE and three TOAs from the BS-UE, BS-HRIS-UE, and BS-HRIS links, and eleven unknown states estimate: the 3D UE position, the 3D HRIS position, and the 3D HRIS orientation, leading to an unsolvable estimation problem. To treat this infeasibility situation, more anchor nodes need to added~\cite{keykhosravi2021semi}. For example, at least two single-antenna BSs will make the scenario feasible, which results in a different setup
.} The element in each $r$th row ($r \in \{0, \dots , M_{_\text{B}}^r- 1\}$) and $s$th column ($s \in \{0, \dots , M_{_\text{B}}^c- 1\}$) has the position $\mathbf{q}_{r,s} =[d(2r-M_{_\text{B}}^r+1)/2), 0, d(2s-M_{_\text{B}}^c+1)/2)]^\top$ in the local coordinate system of the BS, with $d$ being the spacing between the elements. Similarly, the HRIS is a UPA with $M_{_\text{R}} = M_{_\text{R}}^r \times M_{_\text{R}}^c$ unit elements, all attached via a dedicated waveguide to a single RX RF chain \cite{alexandropoulos2021hybrid}, enabling simultaneous tunable reflection and sensing of impinging signals. Accordingly, each $(n,m)$th element ($n \in \{0, \dots , M_{_\text{R}}^r- 1\}$ and $m \in \{0, \dots , M_{_\text{R}}^c- 1\}$) of the HRIS has the position  $\mathbf{q}_{n,m} = [d(2n-M_{_\text{R}}^r+1)/2), 0, d(2m-M_{_\text{R}}^c+1)/2)]^\top$ in the local coordinate system of the HRIS. We assume for both the BS and HRIS that 
$d\leq \lambda/2$ with $\lambda$ being the carrier frequency wavelength. We further assume that the UE operates in the FF range of the BS and the HRIS. Our final assumption is that the HRIS and the UE share their observations with a fusion center (FC) via a reliable link~\cite{Tsinghua_RIS_Tutorial}.  The FC is responsible for carrying out, the targeted in this paper, joint location estimation of the UE and the HRIS, which we will be presented in the sequel.  
\subsection{Signal and Channel Models}\label{sec:sig_model}
The HRIS is capable of simultaneously sensing and reflecting its impinging signals. To this end, it possesses $M_{_\text{R}}$ identical power splitters within its structure to divide the received signal power at each meta-atom into two parts~\cite{alexandropoulos2021hybrid,alamzadeh2021reconfigurable,zhang2022channel}: one for reflection and the other for sensing/reception. For the latter operation, the HRIS adopts a phase shifting network to feed a portion of the impinging signals on its elements to the single RX RF chain. Mathematically, this phase shifting network applies a combining vector modeled by $\mathbf{c}_{_t}\in \mathbb{C}^{M_{_\text{R}}}$ with $|[\mathbf{c}_{_t}]_i| = 1$ $\forall$ $i=1,\ldots,M_{_\text{R}}$ during each time interval $t$. Each $i$-th HRIS unit element also reflects the remaining portion of its impinging signal at each time $t$, which is modeled by the phase shift $[\boldsymbol{\gamma}_{_t}]_i$ with  $\boldsymbol{\gamma}_t\in \mathbb{C}^{M_{_\text{R}}}$ and $|[\boldsymbol{\gamma}_{_t}]_i|= 1$. We make the assumation that the FC has complete knowledge about the HRIS combining vector and phase profile. We further assume that the multi-antenna BS applies the \ac{dft} codebook for its beamforming vector $\mathbf{f}_{_t} \in \mathbb{C}^{M_{_\text{B}} }$ at each time instant $t$, and transmits unit-power symbols.  

Under the above assumptions, after cyclic prefix removal and fast Fourier transform (FFT) application, the received signals at the HRIS and the UE at each time instant $t$, which are respectively denoted by the matrices $\mathbf{y}_{_\text{R},_t}\in \mathbb{C}^{K}$ and $\mathbf{y}_{_\text{U},_t}\in \mathbb{C}^{K}$, can be expressed as follows:\footnote{We assume that there are no scatterers in the BS-HRIS link. This holds from the common placement assumption of (H)RISs in strong LOS conditions from the BS, and mainly close to it~\cite{Marco2019,wu2019towards,keykhosravi2022leveraging,RIS_challenges}. We can also envisage scenarios where the BS and the HRIS are elevated while the UE is located close to the ground.}
\begin{subequations}\label{eq:y_rec}
\begin{align}
\label{eq:y_ris}
\mathbf{y}_{_{\text{R},t}}&= g_{_\text{BR}} \sqrt{\rho P_\text{B}} \mathbf{d}(\tau_{_\text{BR}} )\, \mathbf{c}_{_t}^\top \mathbf{a}_{_\text{R}}(\boldsymbol{\phi}_{_\text{RB}})\mathbf{a}^\top_{_\text{B}}(\boldsymbol{\theta}_{_\text{BR}})\mathbf{f}_{_t} +\mathbf{w}_{_\text{R},_t}, \\ 
\label{eq:y_rx}\mathbf{y}_{_{\text{U},t}} &=\mathbf{y}_{_\text{BU},_t} + \mathbf{y}_{_\text{BRU},_t} +\mathbf{y}^s_{_\text{BU},_t}+\mathbf{y}^s_{_\text{BRU},_t}+\mathbf{w}_{_\text{U},_t} ,
\end{align}
\end{subequations}
where $g_{_\text{BR}}$ denotes the unknown complex gain of the BS-HRIS link, $P_\text{B}$ is the BS transmit power, $\rho$ represents the common power splitting ratio 
at each hybrid meta-atom used for the sensing operation, and $\mathbf{w}_{_\text{R},_t}\in \mathbb{C}^{K}$ and $\mathbf{w}_{_\text{U},_t}\in \mathbb{C}^{K}$ indicate the effect of additive thermal noise at the HRIS and the UE, respectively, each modeled as a vector with zero-mean circularly-symmetric independent and identically distributed Gaussian elements with variance $\sigma^2$. The model can be generalized to accommodate unequal power splitting \cite{zhang2022channel}, but such a generalization is not considered in the current work.
In \eqref{eq:y_rx}, $\mathbf{y}_{_\text{BU},_t}\in \mathbb{C}^{K}$ and $\mathbf{y}_{_\text{BRU},_t}\in \mathbb{C}^{K}$ represent the received signals at time $t$ at the UE from the BS through their direct \ac{los} path and through the HRIS, respectively, while $\mathbf{y}^s_{_\text{BU},_t}\in \mathbb{C}^{K}$ and $\mathbf{y}^s_{_\text{BRU},_t}\in \mathbb{C}^{K}$ denote the received interference signals\footnote{To reduce the complexity in the derivation of the targeted location estimator and its \ac{crb}, we ignore the effects of SPs in the environment on the received signals at the HRIS and UE.} at time $t$ at the \ac{ue} in the BS-UE and BS-HRIS-UE paths, respectively. The latter four contributions at the UE received signal in \eqref{eq:y_rx} are given by:
\begin{subequations}\label{eq:y_UE}
    \begin{align}
    \label{eq:y_txrx}
\mathbf{y}_{_\text{BU},_t}=&g_{_\text{BU}} \sqrt{P_\text{B}} \mathbf{d}(\tau_{_\text{BU}} )\mathbf{a}^\top_{_\text{B}}(\boldsymbol{\theta}_{_\text{BU}})\,\mathbf{f}_{_t},
\\ \label{eq:y_txRISrx}
    \mathbf{y}_{_\text{BRU},_t}=& g_{_\text{BRU}}\sqrt{(1-\rho) P_\text{B}} \mathbf{d}(\tau_{_\text{BRU}} )\,\mathbf{a}^\top_{_\text{R}}(\boldsymbol{\theta}_{_\text{RU}})\,\text{diag} (\boldsymbol{\gamma}_{_t} ) \nonumber \\ &\mathbf{a}_{_\text{R}}(\boldsymbol{\phi}_{_\text{RB}})\,\mathbf{a}^\top_{_\text{B}}(\boldsymbol{\theta}_{_\text{BR}})\mathbf{f}_{_t},\\\label{eq:y_txUESp} \mathbf{y}^s_{_\text{BU},_t}=&\sum_{i=0}^{N_{s_0}} g_{_\text{BSU},_i}\sqrt{P_\text{B}} \mathbf{d}(\tau_{_\text{BSU},_i} )\mathbf{a}^\top_{_\text{B}}(\boldsymbol{\theta}_{_\text{BS},_i})\mathbf{f}_{_t},
\\ \label{eq:y_txRISUESp}    \mathbf{y}^s_{_\text{BRU},_t}=&\sum_{j=0}^{N_{s_1}} g_{_\text{BRSU},_j}\sqrt{(1-\rho) P_\text{B}} \mathbf{d}(\tau_{_\text{BRSU},_j})\mathbf{a}^\top_{_\text{R}}(\boldsymbol{\theta}_{_\text{RS},_j})\text{diag} (\boldsymbol{\gamma}_{_t} ) \nonumber \\& \mathbf{a}_{_\text{R}}(\boldsymbol{\phi}_{_\text{RB}})\,\mathbf{a}^\top_{_\text{B}}(\boldsymbol{\theta}_{_\text{BR}})\mathbf{f}_{_t},
\end{align}
\end{subequations}
where $N_{s_0}$ and $N_{s_1}$ indicate the number of SPs in the BS-UE and HRIS-UE links, respectively. In addition, $g_{_\text{BU}}$ and $g_{_\text{BRU}}$ denote the unknown complex gains of the BS-UE and BS-HRIS-UE links, respectively, while $g_{_\text{BSU},_i}$ and $g_{_\text{BRSU},_j}$ are respectively the unknown complex gains of the BS-$\text{SP}_i$-UE  and BS-HRIS-$\text{SP}_j$-UE links. As previously mentioned, the effect of the reflection of the BS-$\text{SP}_j$-HRIS link via the HRIS on the received signal strength at the UE is neglected in \eqref{eq:y_txRISUESp}.  
Finally, $\mathbf{d}(\cdot)$ appearing in~\eqref{eq:y_ris} and~\eqref{eq:y_txrx}--~\eqref{eq:y_txRISUESp} represents the delay steering vector, which is defined as: 
\begin{equation}\label{eq:delay_tau}
    \mathbf{d}(\tau) \triangleq \left[ 1,\,e^{-\jmath 2\pi \Delta_f \tau},\, \dots, \,e^{-\jmath 2\pi (K-1)\,\Delta_f \tau} \right]^\top,
\end{equation}
where $\Delta_f$ denotes the sub-carrier spacing. Under our lack-of-synchronization assumption between the BS and the other network nodes, the propagation delay for all associated links is given by:
\begin{subequations}
\begin{align}
\label{eq:tau}
    \tau_{_\text{BR}} &= \frac{{d}_{_\text{BR}}}{c}+ b_{_\text{R}}, \tau_{_\text{BRU}} = \frac{{d}_{_\text{BR}}+{d}_{_\text{RU}}}{c}+ b_{_\text{U}}, 
 \\ \label{eq:tau_sp0}
    \tau_{_\text{BSU},_i} &= \frac{d_{_\text{BS},_i}+d_{_\text{US},_i}}{c}+ b_{_\text{U}}, \tau_{_\text{BU}} = \frac{{d}_{_\text{BU}}}{c}+ b_{_\text{U}},   
\\ 
 \label{eq:tau_sp}
     \tau_{_\text{BRSU},_j} &= \frac{d_{_\text{BR}}+d_{_\text{RS},_j}+d_{_\text{US},_j}}{c}+ b_{_\text{U}},    
\end{align}
\end{subequations}
where ${d}_{_\text{BR}}\triangleq\Vert\mathbf{p}_{\text{B}}-\mathbf{p}_{\text{R}}\Vert$, ${d}_{_\text{BU}}\triangleq\Vert\mathbf{p}_{\text{B}}-\mathbf{p}_{\text{U}}\Vert$, ${d}_{_\text{RU}}\triangleq\Vert\mathbf{p}_{\text{R}}-\mathbf{p}_{\text{U}}\Vert$, $d_{_\text{BS},_i} \triangleq\Vert\mathbf{p}_{\text{B}}-\mathbf{p}_{i,s}\Vert$, $d_{_\text{RS},_j} \triangleq\Vert\mathbf{p}_{\text{R}}-\mathbf{p}_{j,s}\Vert$, $d_{_\text{US},_j} \triangleq\Vert\mathbf{p}_{\text{U}}-\mathbf{p}_{j,s}\Vert$, and $d_{_\text{US},_i} \triangleq\Vert\mathbf{p}_{\text{U}}-\mathbf{p}_{i,s}\Vert$ with $c$ being the speed of light. $\mathbf{p}_{i,s} $ and $\mathbf{p}_{j,s} $ denote the position of the $i$th and $j$th SP positions in the BS-UE and HRIS-UE links, respectively.

In \eqref{eq:y_rec} and \eqref{eq:y_UE}, $\boldsymbol{\theta}_{_\text{BR}}$ and $\boldsymbol{\theta}_{_\text{BU}}$ respectively represent the AODs from the BS towards the HRIS and the UE, based on the BS local coordinate system. These vectors are respectively along the directions of $\mathbf{q}_{_\text{BR}}$ and $\mathbf{q}_{_\text{BU}}$ in \eqref{eq:direction_BS}, and can be expressed as: 
\begin{subequations}\label{eq:AODs-BS}
\begin{align}\label{eq:AODs-BS-RIS}
\boldsymbol{\theta}_{_\text{BR}}&=[     {\theta}_{_\text{BR}}^{(\text{az})},
{\theta}_{_\text{BR}}^{(\text{el})}
]^\top=[\text{atan2}([\mathbf{q}_{_\text{BR}}]_2,[\mathbf{q}_{_\text{BR}}]_1),
\text{acos}([\mathbf{q}_{_\text{BR}}]_3)]^\top, \\ \label{eq:AODs-BS-UE}
\boldsymbol{\theta}_{_\text{BU}}&=[
{\theta}_{_\text{BU}}^{(\text{az})},
{\theta}_{_\text{BU}}^{(\text{el})}
]^\top= [\text{atan2}([\mathbf{q}_{_\text{BU}}]_2,[\mathbf{q}_{_\text{BU}}]_1),
\text{acos}([\mathbf{q}_{_\text{BU}}]_3)
]^\top. 
\end{align}
\end{subequations}
Similarly, $\boldsymbol{\theta}_{_\text{RU}}$ and $\boldsymbol{\phi}_{_\text{RB}}$ represent respectively the AOD from the HRIS to the UE and the AOA to the HRIS from the BS, both expressed with respect to the HRIS local coordinate system. In fact, $\boldsymbol{\theta}_{_\text{RU}}$ and $\boldsymbol{\phi}_{_\text{RB}}$ are in the directions of $\mathbf{q}_{_\text{RU}}$ and $\mathbf{q}_{_\text{RB}}$ in \eqref{eq:direction_HRIS}, respectively, and are given as follows:
\begin{subequations}\label{eq:AOD_AOAs-RIS}
\begin{align}\label{eq:AOAs-RIS-BS}
\boldsymbol{\phi}_{_\text{RB}}&=     [
{\phi}_{_\text{RB}}^{(\text{az})},
{\phi}_{_\text{RB}}^{(\text{el})}
]^\top=[\text{atan2}([\mathbf{q}_{_\text{RB}}]_2,[\mathbf{q}_{_\text{RB}}]_1),
\text{acos}([\mathbf{q}_{_\text{RB}}]_3)]^\top,\\ \label{eq:AODs-RIS-UE}
\boldsymbol{\theta}_{_\text{RU}}&=     [{\theta}_{_\text{RU}}^{(\text{az})},
{\theta}_{_\text{RU}}^{(\text{el})}]^\top= [
\text{atan2}([\mathbf{q}_{_\text{RU}}]_2,[\mathbf{q}_{_\text{RU}}]_1),
\text{acos}([\mathbf{q}_{_\text{RU}}]_3)
]^\top. 
\end{align}
\end{subequations}
In the same way, 
$\boldsymbol{\theta}_{_\text{RS},_j}$ represents 
the AOD from the HRIS to the $j$th SP, based on the HRIS local coordinate system, while  $\boldsymbol{\theta}_{_\text{BS},_i}$ denotes the AOD from the BS towards the $i$th SP with respect to the BS local coordinate system.

Finally, in \eqref{eq:y_rec} and \eqref{eq:y_UE}, $\mathbf{a}_{_\text{B}}(\cdot) \in \mathbb{C}^{M_{_\text{B}}}$ and $\mathbf{a}_{_\text{R}}(\cdot)\in \mathbb{C}^{M_{_\text{R}}}$ represent the steering vectors at the BS and the HRIS, respectively, and both can be expressed in the form $\mathbf{a}(\boldsymbol{\psi}) \triangleq \mathbf{a}_r(\boldsymbol{\psi})\otimes \mathbf{a}_c(\boldsymbol{\psi})$, where each 
$n$-th element of $\mathbf{a}_r(\cdot)$ and $\mathbf{a}_c(\cdot)$ is defined as:
\begin{subequations} \label{eq:RISreponses}
\begin{align}\label{eq:ar}
 [\mathbf{a}_r (\boldsymbol{\psi})]_n&= e^{-\jmath \frac{2\pi nd}{\lambda} \sin{[\boldsymbol{\psi}]_\text{el}}\cos{[\boldsymbol{\psi}]_\text{az}}} ,
\\\label{eq:ac} 
 [\mathbf{a}_c (\boldsymbol{\psi})]_n&= e^{-\jmath \frac{2\pi nd}{\lambda} \cos{[\boldsymbol{\psi}]_\text{el}}},
 \end{align}
\end{subequations}
where $\boldsymbol{\psi}$ denotes the AOA/AOD at the HRIS or AOD at the BS. In addition, $[\boldsymbol{\psi}]_\text{az}$ and $[\boldsymbol{\psi}]_\text{el}$ are the azimuth and elevation components of $\boldsymbol{\psi}$, respectively.

\section{Fisher Information Analysis}\label{Sec.FIM}
In this section, we present the Fisher information matrix (FIM) of the associated channels' parameters, the HRIS and UE positions, the HRIS rotation matrix, as well as the HRIS and UE synchronization. 

\subsection{FIM of Channel Parameters}
We first concatenate the $t$-th noise-free observation over the $k$-th sub-carrier at the HRIS, $[\boldsymbol{\mu}_{_\text{R},_t}]_k\in \mathbb{C}$, and at the UE, $[\boldsymbol{\mu}_{_\text{U},_t}]_k\in \mathbb{C}$, in the vector $\boldsymbol{\mu}_{_t,_k} \triangleq [[\boldsymbol{\mu}_{_\text{R},_t}]_k, [\boldsymbol{\mu}_{_\text{U},_t}]_k]^\top \in \mathbb{C}^{2}$. Let us define the aggregated vectors $\boldsymbol{\tau} \triangleq [\tau_{_\text{BR}},\tau_{_\text{BU}},\tau_{_\text{BRU}}]^\top$, $\boldsymbol{\theta} \triangleq [\boldsymbol{\theta}_{_\text{BR}},\boldsymbol{\theta}_{_\text{BU}},\boldsymbol{\theta}_{_\text{RU}}, \boldsymbol{\phi}_{_\text{RB}}]^\top$, and $\mathbf{g}\triangleq[\mathbf{g}^\top_{_\text{BR}},\mathbf{g}^\top_{_\text{BU}},\mathbf{g}^\top_{_\text{BRU}}]^\top$, where $\mathbf{g}_{_\text{BR}} \triangleq[\Re(g_{_\text{BR}}),\Im(g_{_\text{BR}})]^\top$, $\mathbf{g}_{_\text{BU}} \triangleq[\Re(g_{_\text{BU}}),\Im(g_{_\text{BU}})]^\top$, and $\mathbf{g}_{_\text{BRU}} \triangleq[\Re(g_{_\text{BRU}}),\Im(g_{_\text{BRU}})]^\top$. Using the latter notations, we introduce the vector with the associated unknown channels parameters: 
\begin{equation}\label{eq:chan_param}
    \boldsymbol{\zeta}=[\underbrace{\boldsymbol{\tau}^\top,\boldsymbol{\theta}^\top}_{\triangleq\boldsymbol{\eta}\in \mathbb{R}^{11}},\mathbf{g}^\top]^\top \in \mathbb{R}^{17}.
\end{equation}
Considering the availability of the stacked observation vector $\boldsymbol{\mu}_{_t,_k}$ at the fusion center, and given~\eqref{eq:y_ris}--\eqref{eq:y_txRISrx} and the Slepian-Bangs formula~\cite[Sec. 3.9]{kay1993fundamentals}, we can write the FIM of $\boldsymbol{\zeta}$, $\mathbf{J}_{\boldsymbol{\zeta}}\in \mathbb{R}^{17\times17}$, as follows:
\begin{equation}
\label{eq:FIM_ch}
\mathbf{J}_{\boldsymbol{\zeta}}\triangleq \frac{2}{\sigma^2} \sum_{t=1}^T\sum_{k=1}^{K} \Re \Bigg\{\frac{ \partial \boldsymbol{\mu}_{_t,_k}}{\partial\boldsymbol{\zeta}} \left( \frac{ \partial \boldsymbol{\mu}_{_t,_k}}{\partial\boldsymbol{\zeta}}\right)^{\mathsf{H}}\Bigg\}.
\end{equation}
The latter expression can be used to obtain the equivalent FIM (EFIM) of the
AOAs, AODs, and TOAs via the formula:
\begin{align}\label{eq:EFIM}
\mathbf{J}_{\boldsymbol{\eta}} \triangleq [[\mathbf{J}_{\boldsymbol{\zeta}}^{-1}]_{1:11,1:11}]^{-1} \quad.
\end{align}

The CRBs corresponding to AOAs, AODs, and TOAs, namely, the angle of arrival error bound (AAEB), angle of departure error bound (ADEB), and time of arrival error bound (TEB) can be respectively obtained using \eqref{eq:EFIM} as follows:
\begin{subequations}\label{eq:Jch}
\begin{align}\label{eq:TEB_BR}
{\sigma}_{{\tau}_\text{i}}&= \sqrt{\mathbb{E} [\left( \tau_{_\text{i}} -\hat{\tau}_{_\text{i}} \right)^2  ]}\geq \text{TEB}_i\triangleq  \sqrt{\left[ \mathbf{J}_{\boldsymbol{\eta}}^{-1}\right]_{j,j}}\quad,\nonumber \\ &\forall (i,j) \in\{ (\text{BR},1) ,(\text{BU},2) (\text{BRU},3)\} ,
\\
\label{eq:ADEB_BR}{\sigma}_{\boldsymbol{\theta}_\text{i}}&=\sqrt{\mathbb{E} [\Vert \boldsymbol{\theta}_{_\text{i}} -\hat{\boldsymbol{\theta}}_{_\text{i}} \Vert^2  ]}\geq \text{ADEB}_i\triangleq  \sqrt{[ \mathbf{J}_{\boldsymbol{\eta}}^{-1}]_{j,j}}\quad , \nonumber \\&\forall (i,j) \in\{ (\text{BR},4:5) ,(\text{BU},6:7), (\text{BRU},8:9)\} ,
\\
\label{eq:AAE_BR}{\sigma}_{\boldsymbol{\phi}_\text{RB}}&=\sqrt{\mathbb{E}\left [\Vert \boldsymbol{\phi}_{_\text{RB}} -\hat{\boldsymbol{\phi}}_{_\text{RB}} \Vert^2  \right]}\geq \text{AAEB}_{_\text{RB}}\triangleq  \sqrt{[ \mathbf{J}_{\boldsymbol{\eta}_{_\text{ch}}}^{-1}]_{10:11,10:11}},
\end{align}
\end{subequations}
where $\hat{\boldsymbol{\tau}}_{_\text{i}}$, $\hat{\boldsymbol{\theta}}_{_\text{i}}$, and $\hat{\boldsymbol{\phi}}_{_\text{RB}}$ are the estimation of the true parameters ${\boldsymbol{\tau}}_{_\text{i}}$, ${\boldsymbol{\theta}}_{_\text{i}}$, and ${\boldsymbol{\phi}}_{_\text{RB}}$, respectively. 

\subsection{FIM of State Parameters}
We commence by expressing the rotation matrix in \eqref{eq:rot-matrix} with respect to its columns, i.e., as $\mathbf{R} = [\mathbf{r}_1, \mathbf{r}_2, \mathbf{r}_3]$ with $\mathbf{r}_1$, $\mathbf{r}_2$, and $\mathbf{r}_3$ being three-dimensional column vectors. Then, we introduce the following state parameters vector:
\begin{equation}\label{eq:state}
    \boldsymbol{\zeta}_s \triangleq [\mathbf{p}_\text{R}^\top,\mathbf{p}_\text{U}^\top, b_{_R}, b_{_U}, \mathbf{r}^\top] \in \mathbb{R}^{17},
\end{equation}
where $\mathbf{r}\triangleq[\mathbf{r}_1^\top, \mathbf{r}_2^\top, \mathbf{r}_3^\top]^\top$.
Given the relationship between the channel and the state parameters, we can derive the FIM of the latter parameters using the transformation matrix $\mathbf{T} \in \mathbb{R}^{11\times 17}$, where the $\ell$-th row and $m$-th column of $\mathbf{T}$ is obtained as $[\mathbf{T}]_{\ell,m} = {\partial [\,\boldsymbol{\eta}]\,_\ell}/{\partial [\,\boldsymbol{\zeta}_{_\text{s}}]\,_m} $\cite[Eq.(3.30)]{kay1993fundamentals}. The elements of $\mathbf{J}_{\boldsymbol{\zeta}}$ and $\mathbf{T}$ are provided in the Appendices~\ref{Sec.J} and~\ref{Sec.T}, respectively. Then, using these matrices and \eqref{eq:EFIM}, we can compute the FIM of the state parameters as follows:
\begin{equation}\label{ea:FIM_s}
    \mathbf{J}_{\boldsymbol{\zeta}_{_\text{s}}} = \mathbf{T}^\top\mathbf{J}_{\boldsymbol{\eta}} \mathbf{T}.
\end{equation}

To derive the error bounds for estimating the state parameters, it is necessary to consider the constraint on $\mathbf{R}$. Therefore, we derive  the \acf{ccrb}~\cite{ollila2008cramer}, which gives the lower bound on the covariance error of the estimate for each unbiased estimator, taking into account the required constraint on $\mathbf{R}$. The orthogonality of this matrix, i.e., $\mathbf{R}^\top \mathbf{R} = \mathbf{I}_3$, imposes the following constraint:
\begin{align}
\label{ea:rot_const}
    \boldsymbol{h}(\mathbf{r}) \triangleq[&\Vert\mathbf{r}_1\Vert^2 -1, \mathbf{r}_2^\top\mathbf{r}_1,\mathbf{r}_3^\top\mathbf{r}_1, \nonumber \\&\Vert\mathbf{r}_2\Vert^2 -1,\mathbf{r}_2^\top\mathbf{r}_3,\Vert\mathbf{r}_3\Vert^2 -1]^\top= \mathbf{0}_6.
\end{align}
We next define the matrix $\boldsymbol{\Phi}\triangleq\text{blkdiag}(\mathbf{I}_8,\frac{1}{\sqrt{2}}\boldsymbol{\Phi}_0)\in \mathbb{R}^{17\times 17}$ including the matrix notation: 
\begin{align}\label{eq:matrix_psi_0}
\boldsymbol{\Phi}_0 \triangleq\begin{bmatrix}
-\mathbf{r}_3 & \mathbf{0}_3 & \mathbf{r}_2\\
\mathbf{0}_3 & -\mathbf{r}_3 & -\mathbf{r}_1\\
\mathbf{r}_1 & \mathbf{r}_2 & \mathbf{0}_3
\end{bmatrix},
\end{align}
which satisfies $\mathbf{G}(\boldsymbol{\zeta})\boldsymbol{\Phi}=0$ where $[\mathbf{G}]_{i,j}=\partial[\boldsymbol{h}(\boldsymbol{\zeta}_s)]_i/\partial[\boldsymbol{\zeta}_s]_j$ $\forall$$i=1,\ldots,17$ and $\forall$$j=1,\ldots,17$~\cite{ollila2008cramer}. Then, the CCRB of the state parameters can be written as:
\begin{equation}\label{eq:CCRB}
\mathbf{C}_{\boldsymbol{\zeta}_\text{s}}= \boldsymbol{\Phi} (\boldsymbol{\Phi}^\top \mathbf{J}_{\boldsymbol{\zeta}_\text{s}}\boldsymbol{\Phi})^{-1}\boldsymbol{\Phi}^\top.
\end{equation}
Using the latter expression, one can write the position error bound (PEB) of the HRIS and UE, the clock bias error bound (CEB) of the HRIS and UE, and the HRIS orientation error bound (OEB) as follows:
 \begin{subequations}
\begin{align}\label{eq:PEB}
 {\sigma}_{\mathbf{p}_\text{i}}&=\sqrt{\mathbb{E}\left [ \Vert\mathbf{p}_{\text{i}} -\hat{\mathbf{p}}_{\text{i}}\Vert^2   \right]}\geq \text{PEB}_{_\text{i}}\triangleq  \sqrt{ [\mathbf{C}_{\boldsymbol{\zeta}_\text{s}}]_{j:j+2,j:j+2}} , \nonumber \\ &\forall (i,j) \in\{ (\text{R},1) ,(\text{U},4) \} ,
\\
\label{eq:OEB}{\sigma}_{_\mathbf{r}}&= \sqrt{\mathbb{E}\left \Vert(\mathbf{r} -\hat{\mathbf{r}}\Vert^2   \right]}\geq \text{OEB}\triangleq  \sqrt{ [\mathbf{C}_{\boldsymbol{\zeta}_\text{s}}]_{9:17,9:17}} \quad, \\
\label{eq:CEB}
{\sigma}_{{b}_\text{i}}&= \sqrt{\mathbb{E} \left(b_{_\text{i}} -\hat{b}_{_\text{i}}\right)^2   }\geq \text{CEB}_{_\text{i}}\triangleq  \sqrt{ [\mathbf{C}_{\boldsymbol{\zeta}_\text{s}}]_{j,j}},\nonumber \\&\forall (i,j) \in\{ (\text{R},6) ,(\text{U},7) \} , \end{align}
\end{subequations}
where $\hat{\mathbf{p}}_{\text{i}}$, $\hat{\mathbf{r}}$, and $\hat{b}_i$ are the estimates of the true parameters ${\mathbf{p}}_{\text{i}}$, ${\mathbf{r}}$, and ${b}_i$, respectively. 

\section{Proposed Parameters Estimator}\label{sec_estimator}
In this section, we develop a multi-stage estimator that exploits the relationships between the states and channel parameters presented in~Section \ref{sec:sig_model}. To this aim, we sequentially estimate the associated channels. The underlying approach in the derivation of the estimator is the maximum likelihood principle, where the refined parameter search is initialized by a coarse search. In this process, the order of the operation is crucial, leading to different procedures for different links.

\subsection{BS-HRIS Channel Parameter Estimation}\label{Sec:BS-HRISch}
Stacking all observations at the HRIS over $T$ time instants into a matrix $\mathbf{Y}_{_\text{R}}\in \mathbb{C}^{K\times T}$ yields:
\begin{align}\label{eq:y_ris_stack}
   \mathbf{Y}_{_\text{R}} &= g_{_\text{BR}} \sqrt{\rho P_\text{B}}  \mathbf{d}(\tau_{_\text{BR}} )\boldsymbol{a}^\top_{_\text{B}}(\boldsymbol{\theta}_{_\text{BR}},\boldsymbol{\phi}_{_\text{RB}})\,\boldsymbol{\Omega}+\mathbf{W}_{_\text{R}}, 
\end{align}
where we used the notations:
\begin{subequations}
\begin{align}\label{eq:a_br}
   \boldsymbol{a}_{_\text{B}}(\boldsymbol{\theta}_{_\text{BR}},\boldsymbol{\phi}_{_\text{RB}})\triangleq \text{vec} (\,\mathbf{a}_{_\text{R}}(\boldsymbol{\phi}_{_\text{RB}})\mathbf{a}^\top_{_\text{B}}(\boldsymbol{\theta}_{_\text{BR}}))\, \in \mathbb{C}^{M_{_\text{B}}M_{_\text{R}}}, \\\label{eq:Omega}
   \boldsymbol{\Omega}\triangleq[\mathbf{f}_1 \otimes \mathbf{c}_1\,\ldots,\,\mathbf{f}_T \otimes \mathbf{c}_T] \in \mathbb{C}^{M_{_\text{B}}M_{_\text{R}}\times T},
\end{align} 
\end{subequations}
and $\mathbf{W}_{_\text{R}}\in \mathbb{C}^{K \times T}$ is the noise matrix at the HRIS's single RX RF chain over all sub-carriers and time slots; this matrix contains zero-mean circularly-symmetric independent and identically distributed Gaussian elements with variance $\sigma^2$. We then estimate the delay $\tau_{_\text{BR}}$ between BS and HRIS using the approach prresented in~\cite[Secs.~IV-A and~IV-B]{keykhosravi2021siso}. The delay of this path can be obtained via solving the following optimization problem:
\begin{align} \label{eq: TOA_estimation}
    \hat\tau_{_\text{BR}} = \arg\max_{\tau_{_\text{BR}}}\Vert \mathbf{d}^{\mathsf{H}}(\tau_{_\text{BR}}) \mathbf{Y}_{_\text{R}}\Vert.
\end{align}
We can solve the problem~\eqref{eq: TOA_estimation} through 1D line search or gradient-based iterative search with an initial point, which can be provided by an FFT-based method~\cite{ghazalian2022bi}.

We next remove the effect of $\tau_{_\text{BR}}$ from $\mathbf{Y}_{_\text{R}}$ by calculating $\mathbf{Y}_{_\text{R}} \odot (\mathbf{d(-\hat{\tau}_{_\text{BR}}}) \mathbf{1}_{_T}^\top)$. Taking the sum over the subcarriers, and after some algebraic manipulations, the following expression is deduced:
\begin{align}\label{eq:z_r}
    \mathbf{z}_{_\text{R}}  &=  K g_{_\text{BR}} \sqrt{\rho P_\text{B}}\boldsymbol{\Omega}^\top \boldsymbol{a}_{_\text{B}}(\boldsymbol{\theta}_{_\text{BR}},\boldsymbol{\phi}_{_\text{RB}})+\mathbf{v}_{_\text{R}},
\end{align}
where $\mathbf{z}_{_\text{R}} \in \mathbb{C}^{T}$ and   $\mathbf{v}_{_\text{R}}\triangleq (\mathbf{W}_{_\text{R}})^\top \odot (\mathbf{1}_{_T}\mathbf{d}(-\hat{\tau}_{_\text{BR}})^\top) \mathbf{1}_{_{K}}\in \mathbb{C}^{T} $. We next use the matrix $\boldsymbol{\Psi} \triangleq \boldsymbol{\Omega}^\top \boldsymbol{A}\in \mathbb{C}^{T\times J}$ with $J\gg T$, where $\boldsymbol{A}\triangleq [\boldsymbol{a}_{_\text{B}}(\overline{\boldsymbol{\theta}}_{_{\text{BR}_1}},\overline{\boldsymbol{\phi}}_{_{\text{RB}_1}}), \dots, \boldsymbol{a}_{_\text{B}}(\overline{\boldsymbol{\theta}}_{_{\text{BR}_J}},\overline{\boldsymbol{\phi}}_{_{\text{RB}_J}})]\in \mathbb{C}^{M_{_\text{B}}M_{_\text{R}}\times J}$ is a dictionary matrix containing combined array response vectors, given by~\eqref{eq:a_br}, with the AOD from the BS to the HRIS and the AOA at the HRIS from the BS pairs $(\overline{\boldsymbol{\theta}}_{_{\text{BR}_j}},\overline{\boldsymbol{\phi}}_{_{\text{RB}_j}})$ $\forall j = 1,2,\dots, J$. Using this matrix definition, we can approximate $\mathbf{z}_{_\text{R}}$ as follows:
\begin{align}\label{eq:z_r_approx}
    \mathbf{z}_{_\text{R}}  \approx  \boldsymbol{\Psi}\mathbf{x}+\mathbf{v}_{_\text{R}}, 
\end{align}
where $\mathbf{x}\in \mathbb{C}^{J}$ is a sparse vector including a single non-zero element that is approximately equal to $K g_{_\text{BR}} \sqrt{\rho P_\text{B}}$. The latter expression enables us to use compressed sensing (CS) methods to estimate $\mathbf{x}$, the AOA at the HRIS, and the AOD from the BS to the HRIS. To this end, we deploy a 
simple grid search in the dictionary,\footnote{The resolution of the grid searches is fine enough to initialize the Newton method searches, and thus off-grid effects do not limit the achievable accuracy. To avoid the increased complexity of very fine grid searches, a hierarchical approach is followed, where a finer search is done around the optimum obtained with a coarse search.}
 which selects the column of $\boldsymbol{\Psi}$ that has the maximum scalar product with $\mathbf{z}_{_\text{R}}$. 
Finally, we refine the angles estimation. Thus, one can apply these estimated angles as an initial guess for the Newton's method to solve the negative maximum likelihood optimization problem,\footnote{This estimation approach has a weak similarity with the Newtonized orthogonal matching pursuit proposed in~\cite{mamandipoor2016newtonized}. However, in that paper, the proposed algorithm was used to jointly estimate TOA, AOA, and the channel gains\cite{han2019efficient, li2020new}, which constitutes a different goals from this paper. The proposed algorithm herein applies a multi-stage estimator where TOA and AOA/AOD are sequentially estimated, followed by refinement using the Newton method. TOA's estimation is particularly performed using the FFT, followed by fractional refinement.} which can be expressed, using~\eqref{eq:z_r}, as follows:
\begin{align}\label{eq:MLE}\left[\hat{\boldsymbol{\theta}}_{_\text{BR}},\hat{\boldsymbol{\phi}}_{_\text{RB}}\right] 
  =& \arg\min_{\boldsymbol{\theta}_{_\text{BR}},\boldsymbol{\phi}_{_\text{RB}}} \Vert \mathbf{z}_{_\text{R}}-K\sqrt{\rho P_\text{B}} \hat{g}_{_\text{BR}}(\boldsymbol{\theta}_{_\text{BR}},\boldsymbol{\phi}_{_\text{RB}})\nonumber \\ &\quad\quad\quad\quad\quad\,\,\boldsymbol{\Omega}^\top \boldsymbol{a}_{_\text{B}}(\boldsymbol{\theta}_{_\text{BR}},\boldsymbol{\phi}_{_\text{RB}})\Vert^2, 
\end{align}
where $\hat{g}_{_\text{BR}}(\boldsymbol{\theta}_{_\text{BR}},\boldsymbol{\phi}_{_\text{RB}}) \triangleq (\boldsymbol{\Omega}^\top \boldsymbol{a}_{_\text{B}}(\boldsymbol{\theta}_{_\text{BR}},\boldsymbol{\phi}_{_\text{RB}})^\dag\mathbf{z}/(K\sqrt{\rho P_\text{B}})$.

\subsection{BS-UE and BS-HRIS-UE Channel Parameter Estimation}\label{Sec-EST_BS-UE}
We first stack all observations at the UE over $T$ time istants in a matrix $\mathbf{Y}_{_\text{U}}\in \mathbb{C}^{K\times T}$, yielding:
\begin{align}\label{eq:stack_obs_UE}
\mathbf{Y}_{_\text{U}}= \mathbf{Y}_{_\text{BU}}+\mathbf{Y}_{_\text{BRU}}+\mathbf{W}_{_\text{U}}.
\end{align}
The BS-UE and BS-HRIS-UE signal components can be separated using an appropriate design of the HRIS phase profile and BS beam-former\cite{keykhosravi2022ris}, as explained below. The FC can combine the signals in such a way that the interference between both components is eliminated in a static scenario, thus facilitating the derivation of an estimator.

To design the HRIS phase profile and the precoder at the BS, we set $T$ to be an even number and define $\boldsymbol{\gamma}_{2t}=-\boldsymbol{\gamma}_{2t+1}$ and $\mathbf{f}_{2t}=\mathbf{f}_{2t+1}$ for $t=0,1,\dots,T/2$. Given this HRIS phase profile, and considering~\eqref{eq:y_txrx} and~\eqref{eq:y_txRISrx}, we can write: 
\begin{align}\label{eq:post_pros_yUE}
    [\mathbf{Y}_{_\text{BU}}]_{:,2t} &= [\mathbf{Y}_{_\text{BU}}]_{:,2t+1}, \\ \label{eq:post_pros_yBRU}
    [\mathbf{Y}_{_\text{BRU}}]_{:,2t} &= -[\mathbf{Y}_{_\text{BRU}}]_{:,2t+1}. 
\end{align}
Using the latter expressions, the FC performs post-processing to calculate the matrices $\mathbf{Z}_{_\text{BU}}\in \mathbb{C}^{K\times T/2}$ and $\mathbf{Z}_{_\text{BRU}}\in \mathbb{C}^{K\times T/2}$ as follows:
\begin{subequations}
  \begin{align} \label{eq:z_bu}
    [\mathbf{Z}_{_\text{BU}}]_{:,2t} &=[\mathbf{Y}_{_\text{U}}]_{:,2t}+[\mathbf{Y}_{_\text{U}}]_{:,2t+1}=2g_{_\text{BU}} \sqrt{P_\text{B}} \mathbf{d}(\,\tau_{_\text{BU}} )\,\mathbf{a}^\top_{_\text{B}}(\boldsymbol{\theta}_{_\text{BU}})\,\nonumber\\
    &\mathbf{f}_{_\text{2t}}+ \underbrace{[\mathbf{W}_{_\text{U}}]_{:,2t}+[\mathbf{W}_{_\text{U}}]_{:,2t+1}}_{[\mathbf{V}_{_\text{U}}]_{:,2t}},
    \end{align} 
    \begin{align}\label{eq:z_bru}
    [\mathbf{Z}_{_\text{BRU}}]_{:,2t}&=[\mathbf{Y}_{_\text{U}}]_{:,2t}-[\mathbf{Y}_{_\text{U}}]_{:,2t+1}=  2g_{_\text{BRU}}\sqrt{(1-\rho) P_\text{B}} \mathbf{d}(\tau_{_\text{BRU}} )\,\nonumber \\&\mathbf{a}^\top_{_\text{R}}(\boldsymbol{\theta}_{_\text{RU}})\,\text{diag} (\boldsymbol{\gamma}_{_\text{2t}} ) \mathbf{a}_{_\text{R}}(\boldsymbol{\phi}_{_\text{RB}})\,\mathbf{a}^\top_{_\text{B}}(\boldsymbol{\theta}_{_\text{BR}})\mathbf{f}_{_\text{2t}}\nonumber \\&+ \underbrace{[\mathbf{W}_{_\text{U}}]_{:,2t}-[\mathbf{W}_{_\text{U}}]_{:,2t+1}}_{[\mathbf{V}_{_\text{U}}]_{:,2t}},
\end{align}  
\end{subequations}
where $\mathbf{V}_{_\text{U}}\in \mathbb{C}^{K\times T/2}$ is the noise matrix at the UE after post-processing, over all sub-carrier and time slots, containing zero-mean
circularly-symmetric independent and identically distributed
Gaussian elements with variance $2\sigma^2$. As can be observed form~\eqref{eq:z_bu} and~\eqref{eq:z_bru}, the matrices $\mathbf{Z}_{_\text{BU}}$ and $\mathbf{Z}_{_\text{BRU}}$ depend only the parameters of the direct and the reflected channels, respectively. Thus, these channels become separated. It is emphasized here is that the proposed HRIS reflection phase profiles do not lead to a waste of resources due to repeating the beams. This holds because, after the post-processing given by~\eqref{eq:z_bu} and \eqref{eq:z_bru}, the signals $\mathbf{Z}{_\text{BU}}$ and $\mathbf{Z}{_\text{BRU}}$ have high signal-to-noise ratio (SNR) compared to the signals $\mathbf{Y}{_\text{BU}}$ and $\mathbf{Y}{_\text{BRU}}$, respectively.

\subsubsection{BS-UE channel estimation}\label{sec:BS-UE channel estimation}
We commence with BS-UE channel estimation using~\eqref{eq:z_bu}. To this end, we rewrite this expression as follows:
 \begin{align} \label{eq:Z_butf}
    \mathbf{Z}_{_\text{BU}}&=2g_{_\text{BU}} \sqrt{P_\text{B}} \mathbf{d}(\,\tau_{_\text{BU}} )\,\mathbf{a}^\top_{_\text{B}}(\boldsymbol{\theta}_{_\text{BU}})\mathbf{F}+\mathbf{V}_{_\text{U}},
\end{align} 
where $\mathbf{F}\triangleq [\mathbf{f}_{_0}, \dots, \mathbf{f}_{_\text{T/2}}]\in \mathbb{C}^{M\times T/2}$. We next follow a similar approach to that in~Sec.~\ref{Sec:BS-HRISch} to estimate $\tau_{_\textbf{BU}}$ (see expression~\eqref{eq: TOA_estimation}). After removing the effect of this TOA and integrating the signals over the $K$ subcarrier frequencies, the following expression
is deduced:
\begin{align} \label{eq:z_but}
    \mathbf{z}_{_\text{BU}}&=2Kg_{_\text{BU}} \sqrt{P_\text{B}}\mathbf{F}^\top\mathbf{a}_{_\text{B}}(\boldsymbol{\theta}_{_\text{BU}})+\mathbf{v}^0_{_\text{U}},
\end{align}
where $\mathbf{v}^0_{_\text{U}}\triangleq  \mathbf{V}_{_\text{U}}^\top \odot (\mathbf{1}_{_\text{T/2}}\mathbf{d}(-\hat{\tau}_{_\text{BU}})^\top) \mathbf{1}_{_\text{K}}$. 
Using~\eqref{eq:z_but}, we
can write the negative maximum likelihood optimization
problem as:
\begin{align}\label{eq:MLE-theta_BU}
\hat{\boldsymbol{\theta}}_{_\text{BU}}\!&=\!\arg\min_{\boldsymbol{\theta}_{_\text{BU}}} \Vert\mathbf{z}_{_\text{BU}} -2K\sqrt{P_\text{B}}\hat{g}_{_\text{BU}}(\,\boldsymbol{\theta}_{_\text{BU}})\, \mathbf{F}^\top \mathbf{a}_{_\text{B}}(\,\boldsymbol{\theta}_{_\text{BU}})\, \Vert^2,
\end{align}
where $\hat{g}_{_\text{BU}}(\,\boldsymbol{\theta}_{_\text{BU}})\,= \big(\mathbf{F}^\top \mathbf{a}_{_\text{B}}(\,\boldsymbol{\theta}_{_\text{BU}})\,\big)^\dag\mathbf{z}_{_\text{BU}}/(2K\sqrt{P_\text{B}})$. To solve this problem, we first apply a coarse 2D search over $[\boldsymbol{\theta}_{_\text{BU}}]_{_\text{el}}$-$[\boldsymbol{\theta}_{_\text{BU}}]_{_\text{az}}$ search space to jointly estimate the elevation and azimuth angles. Then, we refine the coarse estimation by applying its estimated angles as an initial guess for Newton's method. 

\subsubsection{BS-HRIS-UE channel estimation} \label{sec:BS-HRIS-UE channel estimation}
The remaining parameters that need to be estimated for the BS-HRIS-UE channel are $\tau_{_\text{BRU}}$ and $\boldsymbol{\theta_{_\text{RU}}}$. Based on~\eqref{eq:MLE}, we can rewrite~\eqref{eq:z_bru} as:
\begin{align}\label{eq:z_bru_revise}
    [\mathbf{Z}_{_\text{BRU}}]_{:,2t}&=2g_{_\text{BRU}}\sqrt{(1-\rho) P_\text{B}} \mathbf{d}(\tau_{_\text{BRU}} )\mathbf{a}^\top_{_\text{R}}(\boldsymbol{\theta}_{_\text{RU}})\,\mathbf{b}_{_\text{2t}}+ [\mathbf{V}_{_\text{U}}]_{:,2t},
\end{align}
where $\mathbf{b}_{_\text{2t}}\triangleq\text{diag} (\boldsymbol{\gamma}_{_\text{2t}} ) \mathbf{a}_{_\text{R}}(\hat{\boldsymbol{\phi}}_{_\text{RB}})\,\mathbf{a}^\top_{_\text{B}}(\hat{\boldsymbol{\theta}}_{_\text{BR}})\mathbf{f}_{_\text{2t}}\in \mathbb{C}^{M_{_\text{R}}\times 1}$. Using~\eqref{eq:z_bru_revise}, we then stack all observations related to the BS-HRIS-UE link to express $\mathbf{Z}_{_\text{BRU}}$ as follows:
\begin{align}\label{eq:Z_bru_mat}
    \mathbf{Z}_{_\text{BRU}}&=2g_{_\text{BRU}}\sqrt{(1-\rho) P_\text{B}} \mathbf{d}(\tau_{_\text{BRU}} )\mathbf{a}^\top_{_\text{R}}(\boldsymbol{\theta}_{_\text{RU}})\,\mathbf{B}+ \mathbf{V}_{_\text{U}},
\end{align}
where $\mathbf{B}\triangleq[\mathbf{b}_{_0}, \mathbf{b}_{_2}, \dots, \mathbf{b}_{_\text{T/2}}] \in \mathbb{C}^{M_{_\text{R}}\times T/2}$. As before, $\tau_{_\textbf{BRU}}$ can be estimated using \eqref{eq: TOA_estimation}, and then, it can be removed. By integrating the processed signals over the $K$ subcarrier frequencies, one can obtain the vector $\mathbf{z}_{_\text{BRU}} \in \mathbb{C}^{M_{_\text{R}}}$ as:
\begin{align}\label{eq:z_bru_vec_remov}
    \mathbf{z}_{_\text{BRU}}&=g_{_\text{BRU}}\mathbf{B}_0^\top \mathbf{a}_{_\text{R}}(\boldsymbol{\theta}_{_\text{RU}})\,+ \mathbf{v}^1_{_\text{U}},
\end{align}
where $\mathbf{B}_0\triangleq2K\sqrt{(1-\rho) P_\text{B}}\mathbf{B}^\top$ and $\mathbf{v}^1_{_\text{U}}\triangleq\mathbf{V}_{_\text{U}}^\top \odot (\mathbf{1}_{T/2}\mathbf{d}(-\hat{\tau}_{_\text{BRU}})^\top) \mathbf{1}_{_\text{K}}$. To estimate $\boldsymbol{\theta}_{_\text{RU}}$, we formulate the negative maximum likelihood problem as follows:
\begin{align}\label{eq:MLE-theta_RU}
\hat{\boldsymbol{\theta}}_{_\text{RU}}&=  \arg\min_{\boldsymbol{\theta}_{_\text{RU}}} \Vert    \mathbf{z}_{_\text{BRU}}-\hat{g}_{_\text{BRU}}\mathbf{B}_0^\top \mathbf{a}_{_\text{R}}(\boldsymbol{\theta}_{_\text{RU}}) \Vert^2,
\end{align}
where $\hat{g}_{_\text{BRU}}= (\mathbf{B}_0^\top \mathbf{a}_{_\text{R}}(\boldsymbol{\theta}_{_\text{RU}})\,)^\dag\mathbf{z}_{_\text{BRU}}$.
Similar to Sec.~\ref{sec:BS-UE channel estimation}, we first find a coarse estimate of $\boldsymbol{\theta}_{_\text{RU}}$ through a 2D search. Then, we refine the estimation by applying the coarse estimation of $\boldsymbol{\theta}_{_\text{RU}}$ as an initial point in  Newton's method.
\subsection{Estimation of HRIS and UE Position and Clock Bias} \label{Sec_UE_RIS_CLOCK_est}
Exploiting the one-to-one mapping presented in Section~\ref{sec:sig_model} between the channel and state parameters, and specifically expressions~\eqref{eq:tau} and~\eqref{eq:tau_sp0}, we define the following parameter:
\begin{equation}\label{eq:TDOA}
   \hat{{d}}\triangleq{d}_{_\text{BR}}+{d}_{_\text{RU}}-{d}_{_\text{BU}}= c (\hat{\tau}_{_\text{BRU}}-\hat{\tau}_{_\text{BU}}).
\end{equation}
and the following direction vector:
\begin{align}\label{eq:kappa}
   \boldsymbol{\kappa}(\boldsymbol{\psi})&\triangleq\begin{bmatrix}
       \cos{[\boldsymbol{\psi}]_{_\text{az}}} \sin{[\boldsymbol{\psi}}]_{_\text{el}}\\ \sin{[\boldsymbol{\psi}]_{_\text{az}}} \sin{[\boldsymbol{\psi}]_{_\text{el}}}\\ \cos{[\boldsymbol{\psi}]_{_\text{el}}} 
   \end{bmatrix}.
  \end{align}
Using the latter definition and the estimated AOA/AODs at the BS and HRIS, the angles of the BS-HRIS-UE triangle (see Fig.~\ref{fig:Scenario}) can be obtained as:
\begin{subequations}\label{eq:angles}
  \begin{align}
   \label{eq:betta1}   \beta_0&=\text{acos}\Big(\,\boldsymbol{\kappa}^\top(\boldsymbol{\hat{\theta}}_{_\text{RU}})\boldsymbol{\kappa}(\boldsymbol{\hat{\phi}}_{_\text{RB}}) \Big)\,, \\
   \label{eq:betta0}
\beta_1&=\text{acos}\Big(\,\boldsymbol{\kappa}^\top(\boldsymbol{\hat{\theta}}_{_\text{BU}}) \boldsymbol{\kappa}(\boldsymbol{\hat{\theta}}_{_\text{BR}}) \Big)\,,\\
    \label{eq:betta2}
   \beta_2&=\pi-\beta_0-\beta_1 ,
   \end{align}  
\end{subequations}
Capitalizing on~\eqref{eq:TDOA} and~\eqref{eq:angles}, and applying the law of sines in the triangle with edges the BS, HRIS, and UE in Fig.~\ref{fig:Scenario}, the distances from the BS to the other two network nodes are computed as follows:
\begin{align}\label{eq:dbu}
\hat{d}_{_\text{BU}} = \frac{\hat{d} \sin{\beta_0}}{\sin{\beta_2}+ \sin{\beta_1}-\sin{\beta_0}}, \\
\label{eq:dbr} \,\hat{d}_{_\text{BR}} = \frac{\hat{d} \sin{\beta_2}}{\sin{\beta_2}+ \sin{\beta_1}-\sin{\beta_0}}.
\end{align}
Using~\eqref{eq:kappa}, \eqref{eq:dbu}, \eqref{eq:dbr}, and the AODs from the BS to the other nodes, one can estimate the positions of the HRIS and UE as:
 \begin{align}\label{eq:HRIS-postion_est}
 \hat{\mathbf{p}}_{_\text{R}}=\mathbf{p}_{_\text{B}}+\hat{d}_{_\text{BR}}\boldsymbol{\kappa}(\boldsymbol{\hat{\theta}}_{_\text{BR}}), \quad \text{and} \quad     \hat{\mathbf{p}}_{_\text{U}}=\mathbf{p}_{_\text{B}}+\hat{d}_{_\text{BU}}\boldsymbol{\kappa}(\boldsymbol{\hat{\theta}}_{_\text{BU}}).
\end{align}
Finally, using the estimated TOAs and node positions, we respectively estimate the clock bias at the HRIS and UE as:
    \begin{align}\label{eq:hris-clk_est}
        \hat{b}_{_\text{R}} =\hat{\tau}_{_\text{BR}} - \frac{\hat{d}_{_\text{BR}}}{c}, \quad \text{and} \quad \hat{b}_{_\text{U}} =\hat{\tau}_{_\text{BU}} - \frac{\hat{d}_{_\text{BU}}}{c}.
    \end{align}

\subsection{HRIS Rotation Matrix Estimation}\label{Sec_Rot_est}
We rewrite \eqref{eq:direction_HRIS} 
as follows:

\begin{subequations}\label{eq:direction_HRIS_re}
\begin{equation}\label{eq:direction_HRIS-BS_re}
\boldsymbol{\kappa}(\boldsymbol{{\phi}}_{_\text{RB}})= \mathbf{R}^\top (\mathbf{p}_\text{B}-\mathbf{{p}}_\text{R})/\Vert\mathbf{p}_\text{B}-\mathbf{{p}}_\text{R}\Vert,
\end{equation}
\begin{equation}\label{eq:direction_HRIS_UE_re}
\boldsymbol{\kappa}(\boldsymbol{{\theta}}_{_\text{RU}})= \mathbf{R}^\top (\mathbf{{p}}_\text{U}-\mathbf{{p}}_\text{R})/\Vert\mathbf{{p}}_\text{U}-\mathbf{{p}}_\text{R}\Vert.
 \end{equation}
\end{subequations}
If we replace the values of $\boldsymbol{{\phi}}_{_\text{RB}}$, $\boldsymbol{{\theta}}_{_\text{RU}}$, $\mathbf{p}_\text{R}$, and $\mathbf{p}_\text{U}$ with their estimates (i.e., ~\eqref{eq:MLE},~\eqref{eq:MLE-theta_RU},~\eqref{eq:HRIS-postion_est}, and~\eqref{eq:HRIS-postion_est}, respectively), then the least-squares estimate of the HRIS rotation matrix $\mathbf{R}$ can obtained by solving the following optimization problem:
\begin{subequations}\label{eq:R_opt}
    \begin{align}
 \hat{\mathbf{R}}&= \arg\min_{\mathbf{R}}\Vert\mathbf{Q}-\mathbf{R}\boldsymbol{\Theta}\Vert,\\
&\text{subject to} \quad \mathbf{R}^\top \mathbf{R} = \mathbf{I}_3,\, \text{and} \quad \det(\mathbf{R}) = 1, 
\end{align}
\end{subequations}
where we have used the definitions:
 \begin{subequations}
  \begin{align}\label{eq:Theta}
     \boldsymbol{\Theta}&\triangleq \begin{bmatrix}
         \boldsymbol{\kappa}(\boldsymbol{\hat{\phi}}_{_\text{RB}})&\boldsymbol{\kappa}(\boldsymbol{\hat{\theta}}_{_\text{RU}})
     \end{bmatrix},\\ \label{eq:Q}
     \mathbf{Q}&\triangleq\begin{bmatrix}
         \frac{\mathbf{p}_\text{B}-\mathbf{\hat{p}}_\text{R}}{\Vert\mathbf{p}_\text{B}-\mathbf{\hat{p}}_\text{R}\Vert}& \frac{\mathbf{\hat{p}}_\text{U}-\mathbf{\hat{p}}_\text{R}}{\Vert\mathbf{\hat{p}}_\text{U}-\mathbf{\hat{p}}_\text{R}\Vert}
     \end{bmatrix}.
    \end{align} 
 \end{subequations}
The optimization problem \eqref{eq:R_opt} is known as the \emph{orthogonal Procrustes problem}~\cite{hurley1962procrustes}) and its solutions is given by~\cite[eq. (8)]{schonemann1966generalized,eggert1997estimating} as follows:
\begin{align}\label{eq:optimum_R}
 \hat{\mathbf{R}}&= \mathbf{U}_0\begin{bmatrix}
     1&0&0\\0&1&0\\0&0&\det(\mathbf{U}_0\mathbf{U}_1^\top)
 \end{bmatrix}\mathbf{U}_1^\top,
\end{align}
where $\mathbf{U}_0$ and $\mathbf{U}_1$ are the left and right singular vectors of matrix\footnote{The singular value decomposition (SVD) of $\mathbf{Q}\boldsymbol{\Theta}^\top$ is $\mathbf{Q}\boldsymbol{\Theta}^\top= \mathbf{U}_0\boldsymbol{\Sigma}\mathbf{U}_1^\top$, where $\boldsymbol{\Sigma}$ is a diagonal matrix containing the singular values of $\mathbf{Q}\boldsymbol{\Theta}^\top$.} $\mathbf{Q}\boldsymbol{\Theta}^\top$.
{
\subsection{Estimation Complexity Analysis}
In this subsection, we analyze the computational complexity of the proposed estimator, which is dominated by computation of the channel parameters and rotation matrix. In channel parameters' estimation, we need to compute a 2D $N_F$-point FFT for the delay estimation, whose computational cost is
given by $\mathcal{O}(N_F \log(N_F))$. The joint estimation of $\boldsymbol{\theta_{_\text{BR}}}$ and $\boldsymbol{\phi_{_\text{RB}}}$ needs to build the dictionary, which is requires $\mathcal{O}(TM_{_\text{B}}M_{_\text{R}}J )$ operations. In addition, $\mathcal{O}(TJ)$ operations are needed for searching over the dictionary. Note that the coarse and fine search for the joint estimation of $\boldsymbol{\theta_{_\text{BR}}}$ and $\boldsymbol{\phi_{_\text{RB}}}$ have the same computational complexity order. The computational cost of the final refinement of $\boldsymbol{\theta_{_\text{BR}}}$ and $\boldsymbol{\phi_{_\text{RB}}}$, based on~\eqref{eq:MLE}, is given by $\mathcal{O}(TM_{_\text{B}}M_{_\text{R}} I_1 )$, where $I_1$ indicates the number of iterations. 
For the BS-UE channel estimation (i.e., $\hat{\boldsymbol{\theta}}_{_\text{BU}}$), we resort to Jacobi-Anger expansion to simplify the 2D search into 1D search~\cite{wang2019super}, resulting in complexity $\mathcal{O}(T M_{_\text{B}} (2N + 1) r)$~\cite[Appendix C]{ozturk2023ris}, where $2N + 1$ and $r$ respectively indicate the number of terms in the Jacobi-Anger approximation and the searching dimension in the azimuth or elevation angles. It is worthwhile mentioning that, if we applied a simple 2D search, the computational complexity would be $\mathcal{O}(T M_{_\text{B}}r^2)$; this requires much more computational complexity than the Jacobi-Anger approach since $r\gg (2N+1)$. For the refinement of $\hat{\boldsymbol{\theta}}_{_\text{BU}}$ according to~\eqref{eq:MLE-theta_BU}, the computational complexity is $\mathcal{O}(T M_{_\text{B}} (2N + 1) I_2)$ with $I_2$ denoting the number of iterations. Similarly, the coarse and fine estimations of $\boldsymbol{\theta}_{_\text{RU}}$ are given by $\mathcal{O}(T M_{_\text{R}} (2N + 1) r)$ and $\mathcal{O}(T M_{_\text{R}} (2N + 1) I_3)$, respectively, with $I_3$ being the number of iterations. 
Therefore, the computational complexity of the rotation matrix estimation is bounded to $\mathcal{O}(27)$ operations. Putting all above together, the complexity order of the proposed estimator is:
\begin{align}\label{eq:complexity}
    \mathcal{C}_{\text{prop}} \approx& \mathcal{O}(N_F \log(N_F)) +  \mathcal{O}(TM_{_\text{B}}M_{_\text{R}}(J+I_1)) \nonumber
    \\& + \mathcal{O}(T M_{_\text{B}} N I_2)+ \mathcal{O}(T M_{_\text{R}} N (r+I_3)).
\end{align}  
The term $\mathcal{O}(TM_{_\text{B}}M_{_\text{R}}(J+I_1))$ is in most cases dominant. 

\section{Numerical Results and Discussion}\label{Sec_simulation}
In this section, we evaluate the performance of the proposed estimation algorithm. In particular, we compare the \ac{rmse} of the estimated parameters with their corresponding CRBs, as derived in Sec.~\ref{Sec.FIM}. For the RMSE calculations, we have averaged the results over $500$ independent noise realizations. All the reflection phase shifts of the HRIS hybrid meta-atoms have been drawn from the uniform distribution, i.e., $\angle[\boldsymbol{\gamma}_{_t}]_{k} \sim \mathcal{U}[0,2\pi)$ $\forall$$t$ and  $\forall$$k=1,\ldots,M_{_\text{R}}$. For the HRIS sensing combiner $\boldsymbol{c}_t$ and the BS precoder $\boldsymbol{f}_t$ $\forall$$t$, we have used \ac{dft} codebooks. Each channel gain has been modeled as: 
\begin{align}\label{eq:chakomp_gain}
    g_{_i}\triangleq|g_{_i}| e^{-\jmath \phi_{_i}} \quad i\in \{\text{BR},\text{BU},\text{BRU}\},
\end{align}
where $\phi_{_i}\sim \mathcal{U}[0,2\pi)$ and $|g_{_i}|$ follows the model described in \cite[eq. (21)–(23)]{ellingson2021path}. In addition, $g_{_\text{BSU},_i}$ 
and $g_{_\text{BRSU},_i}$ follow the radar equation~\cite{skolnik1980introduction}. The rest of the simulation parameters are summarized in Table~\ref{table: tab1}. 
 \begin{table}[t!]
\caption{The considered simulation parameters.}
\begin{center}
\begin{tabular}{l c c} 
 \hline \hline
 Parameter & Symbol & Value  \\  
 \hline\hline
 Wavelength & $\lambda$ & $1 ~\text{cm}$
\\HRIS/BS element distance  & $d$ & $0.25 ~\text{cm}$
\\Light speed & $c$ & $3\times 10^{8}~ \text{m}/\text{sec}$\\
 Number of subcarriers & $K$& $128$\\
 Number of transmissions & $T$& $100$\\
 Sub-carrier spacing & $\Delta f$& $120~\text{kHz}$\\
 Noise PSD&$N_0$& $-174~\text{dBm/Hz}$\\ RX's noise figure& $n_f$ & $5~\text{dB}$\\ IFFT Size & $N_F$& $4048$\\ UE position& $\mathbf{p}_\text{U}$& $\left[5~ \text{m},2~\text{m},1~\text{m}\right]^\top$  \\BS position& $\mathbf{p}_{\text{B}}$& $\left[0~\text{m},0~\text{m},0~\text{m}\right]^\top$ \\HRIS position& $\mathbf{p}_{\text{R}}$& $\left[2~ \text{m},12~\text{m},3~\text{m}\right]^\top$\\
 RIS orientation angles& $[\alpha, \gamma, \beta]^\top$& $[20^{\circ},10^{\circ},15^{\circ}]^\top$
 \\Number of BS antennas& $M_{_\text{B}}$& $4\times 4$\\Number of HRIS elements& $M_{_\text{R}}$& $16\times 16$\\
 \hline\hline
\end{tabular}
\label{table: tab1}
\end{center}
\end{table}

\subsection{Channel Parameter Estimation}
We first show the performance of the channel parameter estimation routine from Secs.~\ref{Sec:BS-HRISch} and~\ref{Sec-EST_BS-UE} as a function of the transmit power $P_\text{B}$. To this end, we have selected the $P_\text{B}$ range such that the BRU link is not very weak, to avoid failure of the AOD estimation from the HRIS to the UE, which can be a bottleneck for joint HRIS and UE localization. The results are shown in Fig.~\ref{fig:channelparameterEstimation} considering an HRIS, common for all its hybrid meta-atoms, power splitting factor of $\rho=0.5$. In terms of TOA estimation, we observe in  Fig.~\ref{fig: TEB_pt2} that for the considered range of transmit powers, the TEB bounds coincide with the corresponding RMSE values (denoted by $\sigma_i$ with $i\in\{\tau_\text{BR},\tau_\text{BU},\tau_\text{BRU}\}$). Due to the path loss differences, the TOA estimation of the BRU path is the worst, while the TOA estimation of the BR path is the best. This indicates that the BRU path is the bottleneck in TOA estimation, and thus, in positioning. For the estimation of the AOA and AOD shown in Fig.~\ref{fig: angle_pt2}, the AOA at the HRIS performance is the best. We note that the RMSE of $\boldsymbol{\theta}_{_\text{BR}}$ is smaller that the RMSE of the $\boldsymbol{\theta}_{_\text{BU}}$ because the BS-HRIS link has a higher SNR thanks to the beamforming gain at the HRIS in spite of the larger BS-HRIS distance compared to the BS-UE distance. 
As can be also seen, the worst estimation performance is achieved for the AOD from HRIS to the UE. 
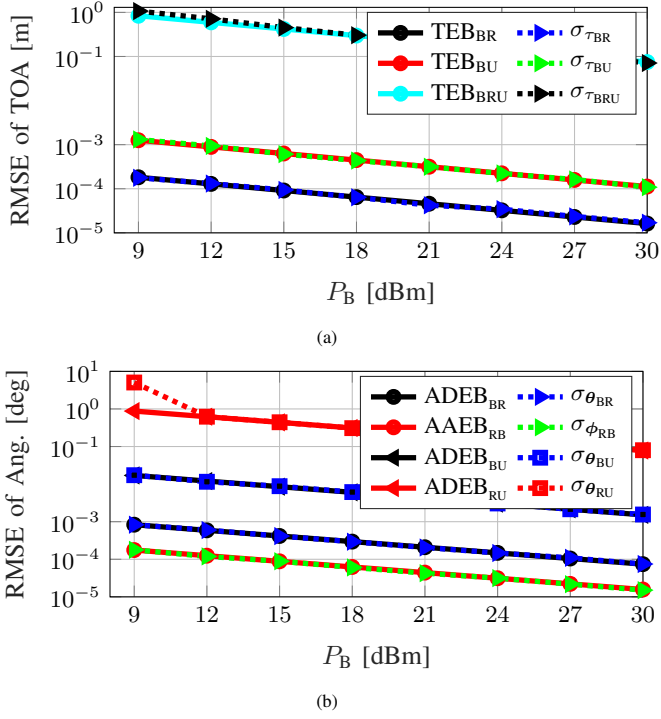
\begin{figure}
\centering
\setlength{\tabcolsep}{0pt}
\begin{tabular}{c}
\subfloat[]{
%
%
\definecolor{mycolor1}{rgb}{0.00000,1.00000,1.00000}%
\begin{tikzpicture}[every node/.append style={font=\small}]

\begin{axis}[%
width=0.8\columnwidth,
height=3cm,
at={(0.758in,0.485in)},
scale only axis,
xmin=8,
xmax=30,
xlabel style={font=\color{white!15!black}},
xlabel={$P_\text{B}$ [dBm]},
ymode=log,
ymin=1e-05,
ymax=1.3,
yminorticks=true,
ylabel style={font=\color{white!15!black}},
ylabel={RMSE of TOA [m]},
axis background/.style={fill=white},
xmajorgrids,
ymajorgrids,
legend columns=2,
legend style={legend cell align=left, align=left, draw=white!15!black}
,legend style={nodes={scale=1, transform shape}},
legend style={legend cell align=left, align=left, draw=white!15!black},
xtick={9, 12,15,18,21,24,27,30},
ytick={10^-5,10^-4,10^-3,10^-1, 10^0}
]
\addplot [color=black, line width=2.0pt, mark=o, mark options={solid, black}]
  table[row sep=crcr]{%
9	0.000182138724454888\\
12	0.00012894434215093\\
15	9.12856034459535e-05\\
18	6.46252581345126e-05\\
21	4.57511790610618e-05\\
24	3.23893543468805e-05\\
27	2.29299068688044e-05\\
30	1.62331309040893e-05\\
};
\addlegendentry{$\text{TEB}_\text{BR}$}

\addplot [color=blue, dotted, line width=2.0pt, mark=triangle, mark options={solid, rotate=270, blue}]
  table[row sep=crcr]{%
9	0.000177165064306075\\
12	0.000131059772305963\\
15	9.39117546543194e-05\\
18	6.39271712069858e-05\\
21	4.223987071447e-05\\
24	3.44132331956819e-05\\
27	2.35908628928561e-05\\
30	1.69633359136059e-05\\
};
\addlegendentry{${\sigma}_{\tau_{\text{BR}}}$}

\addplot [color=red, line width=2.0pt, mark=o, mark options={solid, red}]
  table[row sep=crcr]{%
9	0.00125871284779161\\
12	0.00089110045434416\\
15	0.000630850810115812\\
18	0.000446608171596807\\
21	0.000316174372353431\\
24	0.000223834314037896\\
27	0.00015846255902366\\
30	0.000112182900643517\\
};
\addlegendentry{$\text{TEB}_\text{BU}$}

\addplot [color=green, dotted, line width=2.0pt, mark=triangle, mark options={solid, rotate=270, green}]
  table[row sep=crcr]{%
9	0.00131138624307965\\
12	0.000934618697602105\\
15	0.000609626006755684\\
18	0.000435637108630206\\
21	0.000315106955417975\\
24	0.000222255784824318\\
27	0.000162290556747633\\
30	0.000108482736078787\\
};
\addlegendentry{${\sigma}_{\tau_{\text{BU}}}$}

\addplot [color=mycolor1, line width=2.0pt, mark=o, mark options={solid, mycolor1}]
  table[row sep=crcr]{%
9	0.83908839830591\\
12	0.594029094306344\\
15	0.420540393115645\\
18	0.297719798469497\\
21	0.21076947625414\\
24	0.149213362190983\\
27	0.105634970736888\\
30	0.0747838322167234\\
};
\addlegendentry{$\text{TEB}_\text{BRU}$}

\addplot [color=black, dotted, line width=2.0pt, mark=triangle, mark options={solid, rotate=270, black}]
  table[row sep=crcr]{%
9	1.06995643831119\\
12	0.71435264882015\\
15	0.452121558433473\\
18	0.303460712078499\\
21	0.21255775201697\\
24	0.144070232281543\\
27	0.107872303669917\\
30	0.0710372992288575\\
};
\addlegendentry{${\sigma}_{\tau_{\text{BRU}}}$}

\end{axis}

\end{tikzpicture}%
\label{fig: TEB_pt2}}\\
\subfloat[]{
%
%
\begin{tikzpicture}[every node/.append style={font=\small}]

\begin{axis}[%
width=0.8\columnwidth,
height=3cm,
at={(0.758in,0.485in)},
legend columns=2,
scale only axis,
xmin=8,
xmax=30,
xlabel style={font=\color{white!15!black}},
xlabel={$P_\text{B}$ [dBm]},
ymode=log,
ymin=1e-05,
ymax=10,
yminorticks=true,
ylabel style={font=\color{white!15!black}},
ylabel={RMSE of Ang. [deg]},
axis background/.style={fill=white},
xmajorgrids,
ymajorgrids,
legend style={legend cell align=left, align=left, draw=white!15!black}
,legend style={nodes={scale=1, transform shape}},
legend style={legend cell align=left, align=left, draw=white!15!black},
xtick={9, 12,15,18,21,24,27,30},
ytick={10^-5,10^-4, 10^-3, 10^-1,10^0,10^1}
]
\addplot [color=black, line width=2.0pt, mark=o, mark options={solid, black}]
  table[row sep=crcr]{%
9	0.000831813371572482\\
12	0.000588878769799092\\
15	0.000416894242592689\\
18	0.000295138521577458\\
21	0.000208942072160136\\
24	0.000147919659166228\\
27	0.000104719099134279\\
30	7.41354447766263e-05\\
};
\addlegendentry{$\text{ADEB}_{_\text{BR}}$}

\addplot [color=blue, dotted, line width=2.0pt, mark=triangle, mark options={solid, rotate=270, blue}]
  table[row sep=crcr]{%
9	0.000830739612263544\\
12	0.000598141343342135\\
15	0.000413678230348293\\
18	0.00029303126344621\\
21	0.000205426293619225\\
24	0.000146655900378053\\
27	0.000109290833995458\\
30	7.47250919559138e-05\\
};
\addlegendentry{${\sigma}_{\boldsymbol{\theta}_\text{BR}}$}

\addplot [color=red, line width=2.0pt, mark=o, mark options={solid, red}]
  table[row sep=crcr]{%
9	0.000175915880650394\\
12	0.000124538906112647\\
15	8.81667935742627e-05\\
18	6.24173098335629e-05\\
21	4.41880713692716e-05\\
24	3.12827588459422e-05\\
27	2.21464972488871e-05\\
30	1.56785193662259e-05\\
};
\addlegendentry{$\text{AAEB}_{_\text{RB}}$}

\addplot [color=green, dotted, line width=2.0pt, mark=triangle, mark options={solid, rotate=270, green}]
  table[row sep=crcr]{%
9	0.000178380624503575\\
12	0.000118953177299614\\
15	8.84394412969756e-05\\
18	6.05862119352363e-05\\
21	4.26088402280587e-05\\
24	3.14584701656933e-05\\
27	2.17384655625537e-05\\
30	1.50535065877396e-05\\
};
\addlegendentry{${\sigma}_{\boldsymbol{\phi}_\text{RB}}$}

\addplot [color=black, line width=2.0pt, mark=triangle, mark options={solid, rotate=90, black}]
  table[row sep=crcr]{%
9	0.0170237172196367\\
12	0.0120518688401915\\
15	0.00853206973936227\\
18	0.00604024280405292\\
21	0.00427616442978595\\
24	0.00302729258140042\\
27	0.0021431590210996\\
30	0.00151724039425235\\
};
\addlegendentry{$\text{ADEB}_{_\text{BU}}$}

\addplot [color=blue, dotted, line width=2.0pt, mark=square, mark options={solid, blue}]
  table[row sep=crcr]{%
9	0.0172642302359077\\
12	0.0115435905913955\\
15	0.00880443982681467\\
18	0.00613609344158783\\
21	0.00420209959248061\\
24	0.00305255105593233\\
27	0.00214135445455181\\
30	0.0015522302641113\\
};
\addlegendentry{${\sigma}_{\boldsymbol{\theta}_\text{BU}}$}

\addplot [color=red, line width=2.0pt, mark=triangle, mark options={solid, rotate=90, red}]
  table[row sep=crcr]{%
9	0.877092701194808\\
12	0.620934080324955\\
15	0.439587664546488\\
18	0.311204233982967\\
21	0.220315725530732\\
24	0.155971589123011\\
27	0.110419429003328\\
30	0.0781709692770164\\
};
\addlegendentry{$\text{ADEB}_{_\text{RU}}$}

\addplot [color=red, dotted, line width=2.0pt, mark=square, mark options={solid, red}]
  table[row sep=crcr]{%
9	5.05180399234465\\
12	0.622151712816593\\
15	0.442452772648372\\
18	0.308517084311037\\
21	0.222535087242676\\
24	0.153725021620109\\
27	0.11410117669418\\
30	0.0797805804461741\\
};
\addlegendentry{${\sigma}_{\boldsymbol{\theta}_\text{RU}}$}

\end{axis}

\end{tikzpicture}%
\label{fig: angle_pt2}}
\end{tabular}
\caption{The evaluation of the proposed channel parameter estimator. The power splitting factor is set to $\rho=0.5$. (a)  The RMSE of the TOAs; (c) The RMSE of the AOA/AODs.} \label{fig:channelparameterEstimation}
\end{figure}

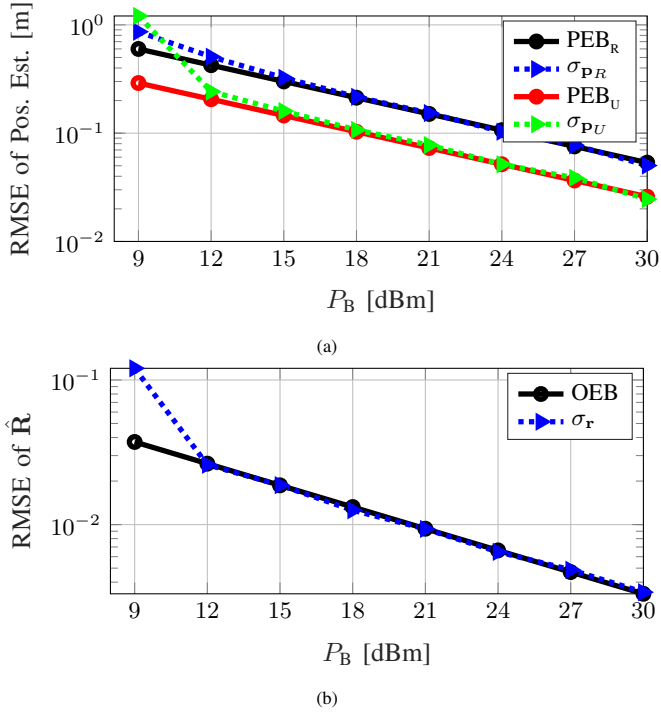
\begin{figure}
\centering
\setlength{\tabcolsep}{0pt}
\begin{tabular}{c c}
\subfloat[]{
%
%
\begin{tikzpicture}[every node/.append style={font=\small}]

\begin{axis}[%
width=0.8\columnwidth,
height=3cm,
at={(0.758in,0.485in)},
scale only axis,
xmin=8,
xmax=30,
xlabel style={font=\color{white!15!black}},
xlabel={$P_\text{B}$ [dBm]},
ymode=log,
ymin=0.01,
ymax=1.20854757367607,
yminorticks=true,
ylabel style={font=\color{white!15!black}},
ylabel={RMSE of Pos. Est. [m]},
axis background/.style={fill=white},
xmajorgrids,
ymajorgrids,
legend style={legend cell align=left, align=left, draw=white!15!black}
,legend style={nodes={scale=1.0, transform shape}},
legend style={legend cell align=left, align=left, draw=white!15!black},
xtick={9, 12,15,18,21,24,27,30},
]
\addplot [color=black, line width=2.0pt, mark=o, mark options={solid, black}]
  table[row sep=crcr]{%
9	0.599465117501201\\
12	0.424388796836821\\
15	0.300444273033813\\
18	0.212698245102697\\
21	0.150578830026234\\
24	0.106601644699237\\
27	0.0754681868171501\\
30	0.0534273858842446\\
};
\addlegendentry{$\text{PEB}_{_\text{R}}$}

\addplot [color=blue, dotted, line width=2.0pt, mark=triangle, mark options={solid, rotate=270, blue}]
  table[row sep=crcr]{%
9	0.864044016464113\\
12	0.507823811177847\\
15	0.324267357000985\\
18	0.21647594758159\\
21	0.15338521731305\\
24	0.103190769975106\\
27	0.0774325763395522\\
30	0.0502431183286953\\
};
\addlegendentry{${\sigma}_{\mathbf{p}_R}$}

\addplot [color=red, line width=2.0pt, mark=o, mark options={solid, red}]
  table[row sep=crcr]{%
9	0.290467550879807\\
12	0.20563527542007\\
15	0.145578632134587\\
18	0.103061773925365\\
21	0.0729621501900766\\
24	0.0516532452543922\\
27	0.0365676980078398\\
30	0.0258879481662263\\
};
\addlegendentry{$\text{PEB}_{_\text{U}}$}

\addplot [color=green, dotted, line width=2.0pt, mark=triangle, mark options={solid, rotate=270, green}]
  table[row sep=crcr]{%
9	1.20854757367607\\
12	0.242084230236581\\
15	0.159959719413038\\
18	0.10715123245718\\
21	0.0775825431132795\\
24	0.051574799920893\\
27	0.0386852375587919\\
30	0.0246398709744189\\
};
\addlegendentry{${\sigma}_{\mathbf{p}_U}$}

\end{axis}

\end{tikzpicture}%
\label{fig: PEB_pt}}\\
\subfloat[]{
%
%
\begin{tikzpicture}[every node/.append style={font=\small}]

\begin{axis}[%
width=0.8\columnwidth,
height=3cm,
at={(0.758in,0.485in)},
scale only axis,
xmin=8,
xmax=30,
xlabel style={font=\color{white!15!black}},
xlabel={$P_\text{B}$ [dBm]},
ymode=log,
ymin=0.00332323171735307,
ymax=0.120413181131139,
yminorticks=true,
ylabel style={font=\color{white!15!black}},
ylabel={RMSE of $\hat{\mathbf{R}}$},
axis background/.style={fill=white},
xmajorgrids,
ymajorgrids,
legend style={legend cell align=left, align=left, draw=white!15!black}
,legend style={nodes={scale=1, transform shape}},
legend style={legend cell align=left, align=left, draw=white!15!black},
xtick={9, 12,15,18,21,24,27,30},
]
\addplot [color=black, line width=2.0pt, mark=o, mark options={solid, black}]
  table[row sep=crcr]{%
9	0.0372872753451642\\
12	0.0263973681207653\\
15	0.0186879057180297\\
18	0.0132300244187735\\
21	0.00936613994282332\\
24	0.00663071917684707\\
27	0.00469418972897327\\
30	0.00332323171735307\\
};
\addlegendentry{OEB}

\addplot [color=blue, dotted, line width=2.0pt, mark=triangle, mark options={solid, rotate=270, blue}]
  table[row sep=crcr]{%
9	0.120413181131139\\
12	0.0259341620687038\\
15	0.018746064648567\\
18	0.0125637400136458\\
21	0.00933377733574122\\
24	0.00647389314780299\\
27	0.00487446116280827\\
30	0.00340803785543445\\
};
\addlegendentry{${\sigma}_{\mathbf{r}}$}

\end{axis}

\end{tikzpicture}%
\label{fig: Ori_pt}}
\end{tabular}
\caption{The evaluation of the proposed UE and RIS state estimator. The power splitting factor is set to $\rho=0.5$. (a) The RMSE of the position estimations; (b) The RMSE of the HRIS's rotation matrix estimation.} \label{fig:localization}
\end{figure}

\subsection{UE and HRIS Position Estimation}
    In Fig.~\ref{fig:localization}, we present numerical results for the UE and HRIS positioning performance, as well as the HRIS orientation estimation performance. As depicted in Fig.~\ref{fig: PEB_pt}, the UE can be localized somewhat better than the HRIS. However, the positioning performance difference is small, since the two spatial states are coupled. 
In terms of the HRIS orientation estimation, as illustrated in Fig.~\ref{fig: Ori_pt}, the proposed estimator achieves the bound for most considered transmit power levels. Nevertheless, the estimations for the HRIS and the UE positions and that for the HRIS orientation affect each other's accuracy.\footnote{The role of passive RISs in localizing UEs has been studied in the literature, e.g., \cite{keykhosravi2021siso,Multiple_passive_RIS}. In this paper, we focus on the joint estimation of the HRIS state and UE position, and leave the investigation on the UE localization accuracy improvement with an HRIS, as compared to a solely reflective RIS~\cite{Tsinghua_RIS_Tutorial}, for a future work.} To investigate this fact, we consider the scenarios described in Table~\ref{table: tab2}. 
\begin{table}
\caption{The considered scenarios for studying the effect of the HRIS state on the UE state estimation, and vice versa.}
\begin{center}
\begin{tabular}{l c  c c } 
 \hline \hline
 Scenarios & $\mathbf{p}_{\text{R}}$ & $\mathbf{p}_{\text{U}}$& $\mathbf{R}$  \\  
 \hline\hline
 C1 & unknown & unknown & unknown 
\\  \hline
C2& unknown & known & known 
\\ 
 \hline
 C3& unknown & unknown & known 
\\ \hline 
C4& unknown & known & unknown 
\\ \hline
C5& known & unknown & known 
\\ \hline
C6& known & unknown & unknown 
\\ 
 \hline\hline
\end{tabular}
\label{table: tab2}
\end{center}
\end{table}
Based on these scenarios, we calculate the CRBs on the HRIS and UE position estimations, which are shown in Fig.~\ref{fig:ris_ue_pos_eff}. As it can be observed, having information about the UE position causes the improvement in the HRIS positioning error (compare C1 with C2 and C4), while the HRIS orientation does not have a considerable impact (compare C1 with C3). Similarly, the UE position estimation comes with smaller error, i.e., compare C1 with C5 and C6, when the HRIS position is known, as compared with the unknown HRIS position case, i.e., compare C1 with C3. One can also see from comparing C1 and C3 in Fig.~\ref{fig:ris_ue_pos_eff} that the PEB of the UE and the HRIS are the same regardless of whether the state of the matrix $\mathbf{R}$ is known or not.  
This happen because matrix $\mathbf{R}$ is not used in the estimation of the UE and HRIS positions (see Sec.~\ref{Sec_UE_RIS_CLOCK_est}). All parameters except the HRIS rotation matrix are estimated based on solving the triangle formed by the BS, HRIS, and UE, and the relative angle between the BS-HRIS AOA and the HRIS-UE AOD is independent of having prior knowledge of the HRIS rotation. On the contrary, the estimation of the HRIS orientation uses the UE and HRIS estimated positions (see Sec.~\ref{Sec_Rot_est}).
\begin{figure}
\centering
\includegraphics[width=0.8\linewidth]{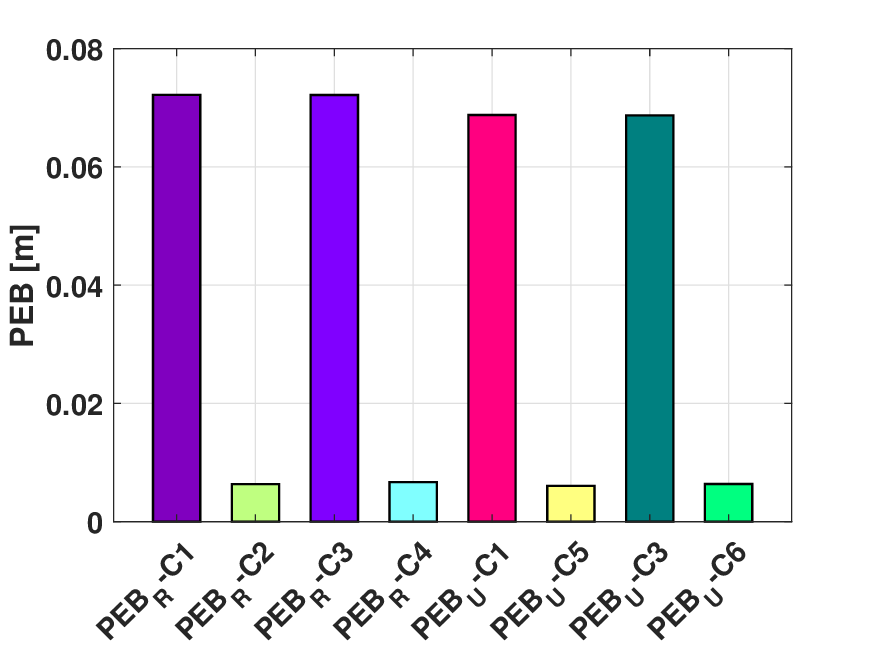}
\caption{The role HRIS position estimation on the UE positioning accuracy, and vice versa. The HRIS common power splitting factor and the BS transmit power were set to $\rho=0.5$ and $P_\text{B} = 15~\text{dBm}$, respectively.}
\label{fig:ris_ue_pos_eff}
\end{figure}
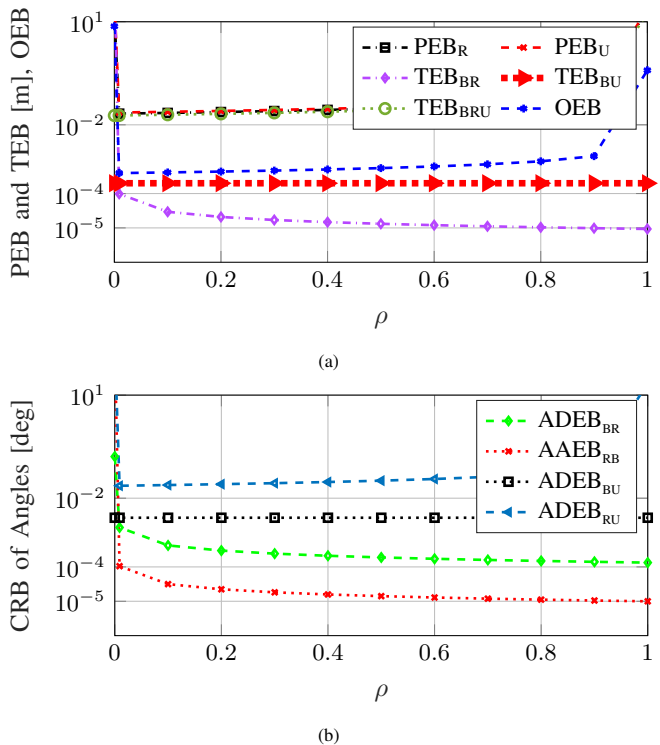
\begin{figure}
\centering
\setlength{\tabcolsep}{0pt}
\begin{tabular}{c c}
\subfloat[]{
%
%
\definecolor{mycolor1}{rgb}{0.71765,0.27451,1.00000}%
\definecolor{mycolor2}{rgb}{0.46667,0.67451,0.18824}%
\begin{tikzpicture}[every node/.append style={font=\small}]

\begin{axis}[%
width=0.8\columnwidth,
height=3.2cm,
at={(0.758in,0.485in)},
legend columns = 2,
scale only axis,
xmin=0,
xmax=1,
xlabel style={font=\color{white!15!black}},
xlabel={$\rho$},
ymode=log,
ymin=1e-06,
ymax=10,
ylabel style={font=\color{white!15!black}},
ylabel={PEB and TEB [m], OEB},
axis background/.style={fill=white},
xmajorgrids,
ymajorgrids,
legend style={legend cell align=left, align=left, draw=white!15!black}
,legend style={nodes={scale=1.0, transform shape}},
xtick={0,0.2,0.4,0.6,0.8,1},
ytick={10^-5,10^-4,10^-2,10^1}
]
\addplot [color=black, dashdotted, line width=1.0pt, mark size=1.5pt, mark=square, mark options={solid, black}]
  table[row sep=crcr]{%
1e-15	15.7286410462582\\
0.009	0.0211495582685561\\
0.1	0.0221925685770949\\
0.2	0.0235385846779735\\
0.3	0.0251636154205704\\
0.4	0.0271796102606491\\
0.5	0.0297735516085121\\
0.6	0.0332875957132417\\
0.7	0.0384369226436179\\
0.8	0.0470750651145193\\
0.9	0.0665737340899609\\
0.999999	22.4116987831458\\
};
\addlegendentry{$\text{PEB}_{\text{R}}$}

\addplot [color=red, dashed, line width=1.0pt, mark size=1.5pt, mark=x, mark options={solid, red}]
  table[row sep=crcr]{%
1e-15	46.0534737562052\\
0.009	0.022803839020243\\
0.1	0.0239286788668204\\
0.2	0.0253798935932466\\
0.3	0.0271319233655571\\
0.4	0.0293054778287011\\
0.5	0.0321021592131792\\
0.6	0.0358908742481158\\
0.7	0.0414427170665702\\
0.8	0.0507561344827098\\
0.9	0.0717791848949645\\
0.999999	24.1586965224237\\
};
\addlegendentry{$\text{PEB}_{\text{U}}$}

\addplot [color=mycolor1, dashdotted, line width=1.0pt, mark size=1.5pt, mark=diamond, mark options={solid, mycolor1}]
  table[row sep=crcr]{%
1e-15	294.414464005537\\
0.009	9.813815466853e-05\\
0.1	2.94414464005497e-05\\
0.2	2.08182463977689e-05\\
0.3	1.69980270046935e-05\\
0.4	1.47207232002748e-05\\
0.5	1.31666151015093e-05\\
0.6	1.20194201618108e-05\\
0.7	1.11278207734177e-05\\
0.8	1.04091231988845e-05\\
0.9	9.81381546684896e-06\\
0.999999	9.31020747865719e-06\\
};
\addlegendentry{$\text{TEB}_\text{BR}$}

\addplot [color=red, dotted, line width=2.5pt, mark size=1.5pt, mark=triangle, mark options={solid, rotate=270, red}]
  table[row sep=crcr]{%
1e-15	0.000199492542372852\\
0.009	0.000199492542372852\\
0.1	0.000199492542372852\\
0.2	0.000199492542372852\\
0.3	0.000199492542372852\\
0.4	0.000199492542372852\\
0.5	0.000199492542372852\\
0.6	0.000199492542372852\\
0.7	0.000199492542372852\\
0.8	0.000199492542372852\\
0.9	0.000199492542372852\\
0.999999	0.000199492542372852\\
};
\addlegendentry{$\text{TEB}_\text{BU}$}

\addplot [color=mycolor2, dotted, line width=1.0pt, mark=o, mark options={solid, mycolor2}]
  table[row sep=crcr]{%
1e-15	0.0186540354936951\\
0.009	0.0187385495531083\\
0.1	0.0196630799045663\\
0.2	0.0208558457093006\\
0.3	0.0222958369016081\\
0.4	0.0240822562688814\\
0.5	0.0263807899881728\\
0.6	0.0294946198568536\\
0.7	0.0340574534279966\\
0.8	0.0417116914186011\\
0.9	0.0589892397137072\\
0.999999	18.6540354934294\\
};
\addlegendentry{$\text{TEB}_\text{BRU}$}

\addplot [color=blue, dashed, line width=1.0pt, mark size=1.5pt, mark=asterisk, mark options={solid, blue}]
  table[row sep=crcr]{%
1e-15	7.39452907804718\\
0.009	0.0003934738087179\\
0.1	0.0004114686897858\\
0.2	0.000435958369259549\\
0.3	0.000465595950376672\\
0.4	0.000502411139012109\\
0.5	0.000549831308209602\\
0.6	0.00061413678666952\\
0.7	0.000708460077196594\\
0.8	0.000866845648848881\\
0.9	0.00122472102118966\\
0.999999	0.386978058712559\\
};
\addlegendentry{OEB}

\end{axis}

\end{tikzpicture}%
\label{fig: PEB_rho}}\\
\subfloat[]{
%
%
\definecolor{mycolor1}{rgb}{0.00000,0.44706,0.74118}%
\begin{tikzpicture}[every node/.append style={font=\small}]

\begin{axis}[%
width=0.8\columnwidth,
height=3.2cm,
at={(0.758in,0.485in)},
scale only axis,
xmin=0,
xmax=1,
xlabel style={font=\color{white!15!black}},
xlabel={$\rho$},
ymode=log,
ymin=1e-06,
ymax=10,
yminorticks=true,
ylabel style={font=\color{white!15!black}},
ylabel={CRB of Angles [deg]},
axis background/.style={fill=white},
xmajorgrids,
ymajorgrids,
xlabel style={font=\color{white!15!black}},
legend style={legend cell align=left, align=left, draw=white!15!black}
,legend style={nodes={scale=1.0, transform shape}},
xtick={0,0.2,0.4,0.6,0.8,1},
ytick={10^-5,10^-4,10^-2,10^1}
]
\addplot [color=green, dashed, line width=1.0pt, mark size=1.5pt, mark=diamond, mark options={solid, green}]
  table[row sep=crcr]{%
1e-15	0.162242537539093\\
0.009	0.00140611748209557\\
0.1	0.000421862469272977\\
0.2	0.00029830276481025\\
0.3	0.000243563446668374\\
0.4	0.000210932244450349\\
0.5	0.000188663595109152\\
0.6	0.000172225548014081\\
0.7	0.000159449898279264\\
0.8	0.000149151739430467\\
0.9	0.000140621620969659\\
0.999999	0.00013340545932318\\
};
\addlegendentry{$\text{ADEB}_{_\text{BR}}$}

\addplot [color=red, dotted, line width=1.0pt, mark size=1.5pt, mark=x, mark options={solid, red}]
  table[row sep=crcr]{%
1e-15	285.925014255949\\
0.009	0.000105133447414009\\
0.1	3.15403659164248e-05\\
0.2	2.23024182203751e-05\\
0.3	1.820985138044e-05\\
0.4	1.57701952616602e-05\\
0.5	1.41052921830498e-05\\
0.6	1.28763116279127e-05\\
0.7	1.19211484113159e-05\\
0.8	1.11512134601464e-05\\
0.9	1.05134650267347e-05\\
0.999999	9.97395377776123e-06\\
};
\addlegendentry{$\text{AAEB}_{_\text{RB}}$}

\addplot [color=black, dotted, line width=1.0pt, mark size=1.5pt, mark=square, mark options={solid, black}]
  table[row sep=crcr]{%
1e-15	0.00269807735317844\\
0.009	0.00269807735317844\\
0.1	0.00269807735317844\\
0.2	0.00269807735317844\\
0.3	0.00269807735317844\\
0.4	0.00269807735317844\\
0.5	0.00269807735317844\\
0.6	0.00269807735317844\\
0.7	0.00269807735317844\\
0.8	0.00269807735317844\\
0.9	0.00269807735317844\\
0.999999	0.00269807735317844\\
};
\addlegendentry{$\text{ADEB}_{_\text{BU}}$}

\addplot [color=mycolor1, dashed, line width=1.0pt, mark size=1.5pt, mark=triangle, mark options={solid, rotate=90, mycolor1}]
  table[row sep=crcr]{%
1e-15	296.921776978168\\
0.009	0.0229960566404886\\
0.1	0.0241303814011671\\
0.2	0.0255941206218243\\
0.3	0.0273612609130938\\
0.4	0.0295535368591269\\
0.5	0.0323742758304029\\
0.6	0.0361955394327925\\
0.7	0.0417950077921758\\
0.8	0.0511882204532145\\
0.9	0.0723910745097113\\
0.999999	22.892067495948\\
};
\addlegendentry{$\text{ADEB}_{_\text{RU}}$}

\end{axis}

\end{tikzpicture}%
\label{fig: ang_rho}}
\end{tabular}
\caption{
The effect of the HRIS common power splitting ration $\rho$ on the estimation performance when the BS transmit power is set to $P_\text{B}=\text{25dBm}$. (a) The effect on the position, TOA, and rotation matrix estimations; (b) The effect on the angles estimations.}
\label{fig:rho_effect}
\end{figure}
\begin{figure}
\centering
\setlength{\tabcolsep}{0pt}
\begin{tabular}{c c}
\subfloat[]{\includegraphics[width=.9\linewidth]{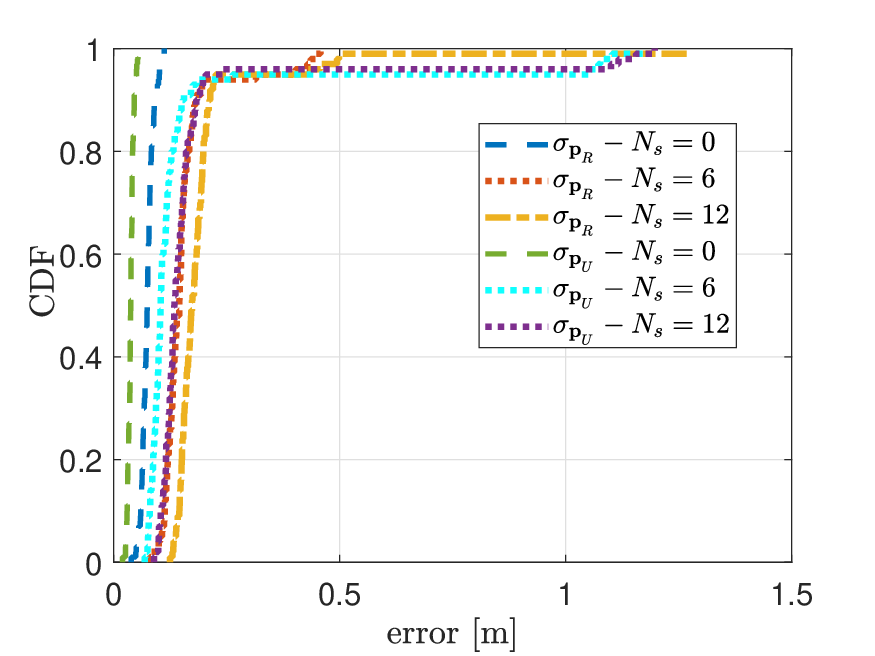}
\label{fig: Pos_esp}}\\
\subfloat[]{\includegraphics[width=.9\linewidth]{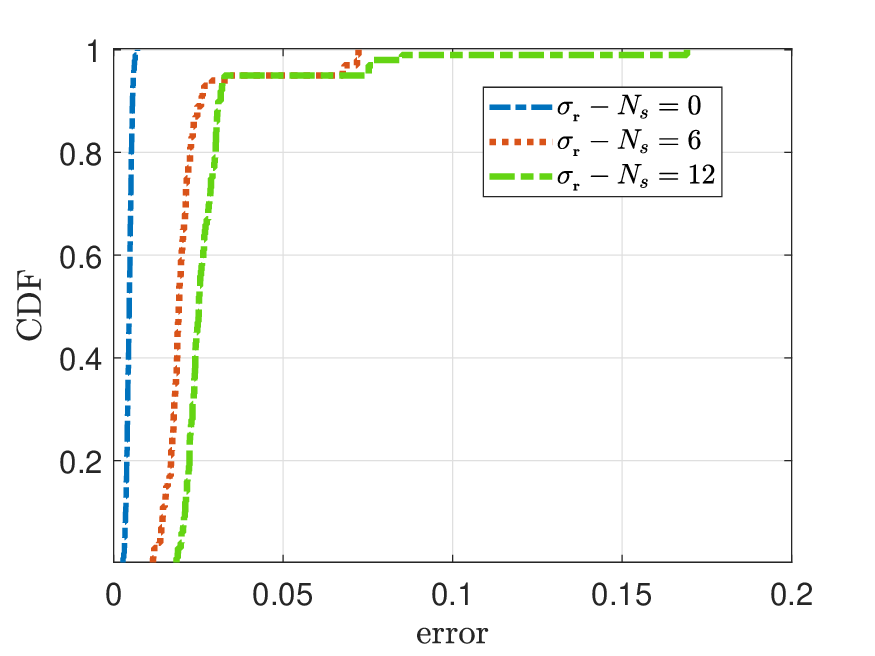}
\label{fig: Ori_sp}}
\end{tabular}
\caption{The evaluation of the proposed estimator for the case where SPs exist in the BS-UE and HRIS-UE links, which are randomly distributed in $[x_0,y_0,z_0]$ and $[x_1,y_1,z_1]$, with $x_0\sim \mathcal{U}(-8\text{m},8\text{m})$, $y_0\sim \mathcal{U}(0\text{m},3\text{m})$, and $z_0\sim \mathcal{U}(-5\text{m},1\text{m})$, as well as $x_1\sim \mathcal{U}(2.5\text{m},4.5\text{m})$ and $y_1\sim \mathcal{U}(4\text{m},11\text{m})$, $z_1\sim \mathcal{U}(-5\text{m},1\text{m})$. Their radar cross section is $1\text{m}^2$ and the HRIS common power splitting factor is set to $\rho=0.5$. (a) The effect on the estimation of the UE and HRIS positions; and (b) The effect on the estimation of $\mathbf{R}$.}
\label{fig:sp_effect}
\end{figure}


\subsection{The Role of the HRIS Power Splitting Ratio $\rho$}
We investigate the effect of $\rho$ on the estimation accuracy in Figs.~\ref{fig: PEB_rho} and~\ref{fig: ang_rho}. As previously demonstrated, the sensing/reception power at the HRIS improves with increasing $\rho$. Therefore, as it can be observed, when the value of $\rho$ increases, the corresponding CRBs of the BS-HRIS channel parameters' estimation (i.e., $\text{ADEB}_{_\text{BR}}$, $\text{AAEB}_{_\text{RB}}$, and $\text{TEB}_{_\text{BR}}$) decline. Furthermore, we can see that the CRB of the HRIS-UE channel parameters exhibits a high error variance as $\rho$ increases. 
We also observe that, as $\rho$ tends to zero, the signal processed at the HRIS becomes very weak, and the error in the estimation of the BS-HRIS delay, AOD, and AOA increases. As expected, the PEB and OEB of the UE degrade as well. Conversely, when $\rho$ tends to one, the signal received by the UE via the BS-HRIS-UE link becomes very weak, and the error in the corresponding delay and AOD increases, again affecting the PEB and OEB of the UE. Therefore, there is a value of $\rho$ that minimizes the PEB and the OEB, which is not necessarily the same for both metrics. Roughly speaking a value around $\rho=0.5$ leads to a reasonable performance trade-off; similarly has been observed for full channel state information estimation~\cite{alexandropoulos2021hybrid}.

\subsection{The Impact of SPs on the Estimation Performance}
We finally investigate the robustness of the proposed estimation approach in the presence of additional SPs in the wireless environment. Figures~\ref{fig: Pos_esp} and~\ref{fig: Ori_sp} illustrate the estimation error of the state parameters for $20$ realizations of the SP positions. Without loss of generality, we set 
$N_{s_0}=N_{s_1}$, and we use the notation $N_s\triangleq N_{s_0}+N_{s_1}$. The SPs in the BS-UE link are randomly distributed in the channel environment with coordinates $[x_0,y_0,z_0]$, where $x_0\sim \mathcal{U}(-8\text{m},8\text{m})$, $y_0\sim \mathcal{U}(0\text{m},3\text{m})$, and $z_0\sim \mathcal{U}(-5\text{m},1\text{m})$.  Similarly, the SPs in the HRIS-UE link are placed randomly in the environment with coordinates $[x_1,y_1,z_1]$, where $x_1\sim \mathcal{U}(2.5\text{m},4.5\text{m})$, $y_1\sim \mathcal{U}(4\text{m},11\text{m})$, and $z_1\sim \mathcal{U}(-5\text{m},1\text{m})$. We model the channel gain for the SPs according to the radar equation~\cite{skolnik1980introduction}, considering that the radar cross-section is $1 \mathrm{m}^2$. As can be seen from both figures, the interference from the SPs deteriorates the estimation accuracy. It is shown that this accuracy degrades as the number $N_s$ of SPs increases. Interestingly, it is depicted that the proposed estimation performs satisfactorily even in the presence of a large number of SPs, but is affected by a small number of relatively large outliers.

\section{Conclusions}\label{Sec_conclusion}
In this paper, we presented a multi-stage estimator for the unknown 3D rotation matrix and 3D position of a single-RX-RF HRIS and the unknown 3D position of a single-antenna UE in a multi-carrier system with a multi-antenna BS. The proposed estimation approach exploits the geometrical channel features to effectively estimate the unknown state parameters. Our simulation results confirmed the validity of the presented approach and showcased that the RMSE of the estimations attains the corresponding CRBs within a certain transmit power range. Moreover, it was demonstrated that the \ac{hris} power splitting ratio between the reflecting and sensing operations of the hybrid meta-atoms plays a critical role on the overall estimation accuracy. It was finally demonstrated that the proposed joint 3D user and 6D HRIS localization method is robust enough when additional \acp{SP} are present in the wireless  propagation environment under investigation.

\appendices 
\section{Derivation of $\mathbf{J}_{\boldsymbol{\zeta}}$}\label{Sec.J}
To calculate $\mathbf{J}_{\boldsymbol{\zeta}}$ in \eqref{eq:FIM_ch}, we need to compute $\partial\boldsymbol{\mu}_{_t,_k}/\partial\boldsymbol{\zeta}=\left[{ \partial [\boldsymbol{\mu}_{_\text{R},_t}]_k}/{\partial\boldsymbol{\zeta}},{ \partial [\boldsymbol{\mu}_{_\text{U},_t}]_k}/{\partial\boldsymbol{\zeta}} \right]^\top\in \mathbb{C}^{17}$.
By considering the noise-free parts of the observations at the HRIS and the UE, as given by~\eqref{eq:y_rec} and \eqref{eq:y_UE}, we can derive the following expressions:
\begin{subequations}
\allowdisplaybreaks
\label{eq:J_deriv}
\begin{align}\label{eq:mu0_tauBR}
  \frac{ \partial [\boldsymbol{\mu}_{_\text{R},_t}]_k}{\partial{\tau_{_\text{BR}}}}&= g_{_\text{BR}} \sqrt{\rho P_\text{B}} [\mathbf{\dot{d}}(\tau_{_\text{BR}})]_k \mathbf{c}_{_t}^\top \mathbf{a}_{_\text{R}}(\boldsymbol{\phi}_{_\text{RB}})\mathbf{a}^\top_{_\text{B}}(\boldsymbol{\theta}_{_\text{BR}})\mathbf{f}_{_t}, 
\\\label{eq:mu1_tauBR_and mu0_tauBU}  \frac{ \partial [\boldsymbol{\mu}_{_\text{U},_t}]_k}{\partial{\tau_{_\text{BR}}}} &= 0, \quad\frac{ \partial [\boldsymbol{\mu}_{_\text{R},_t}]_k}{\partial{\tau_{_\text{BU}}}} = 0, \quad
  \frac{ \partial [\boldsymbol{\mu}_{_\text{R},_t}]_k}{\partial{\tau_{_\text{BRU}}}} = 0,
\\\label{eq:mu1_tauBU}  \frac{ \partial [\boldsymbol{\mu}_{_\text{U},_t}]_k}{\partial{\tau_{_\text{BU}}}}&= g_{_\text{BU}} \sqrt{P_\text{B}}   [\mathbf{\dot{d}}(\,\tau_{_\text{BU}} )\,]_{_{k}}\mathbf{a}^\top_{_\text{B}}(\,\boldsymbol{\theta}_{_\text{BU}})\,\mathbf{f}_{_t} , 
\\\label{eq:mu1_tauBRU}  \frac{ \partial [\boldsymbol{\mu}_{_\text{U},_t}]_k}{\partial{\tau_{_\text{BRU}}}} =& g_{_\text{BRU}}\sqrt{(\,1-\rho)\, P_\text{B}} [\mathbf{\dot{d}}(\,\tau_{_\text{BRU}} )\,]_{k}\mathbf{a}_{_\text{R}}^\top(\,\boldsymbol{\theta}_{_\text{RU}})\, \nonumber \\&\text{diag} (\,\boldsymbol{\gamma}_{_t})\, \mathbf{a}_{_\text{R}}(\,\boldsymbol{\phi}_{_\text{RB}})\,\mathbf{a}^\top_{_\text{B}}(\,\boldsymbol{\theta}_{_\text{BR}})\,\mathbf{f}_{_t}, 
\\\label{eq:mu0_thetaBR_el}  \frac{ \partial [\boldsymbol{\mu}_{_\text{R},_t}]_k}{\partial{[\boldsymbol{\theta}_{_\text{BR}}}]_{_i}} =& g_{_\text{BR}} \sqrt{\rho P_\text{B}}   [\mathbf{d}(\,\tau_{_\text{BR}} )\,]_{_{k}}\mathbf{c}_{_t}^\top \mathbf{a}_{_\text{R}}(\,\boldsymbol{\phi}_{_\text{RB}})\,\frac{\partial{\mathbf{a}^\top_{_\text{B}}(\,\boldsymbol{\theta}_{_\text{BR}})\,}}{\partial{[\boldsymbol{\theta}_{_\text{BR}}}]_{_i}}\mathbf{f}_{_t} \nonumber\\&\quad\forall i\in \{\text{el},\text{az}\},  
\\\label{eq:mu1_thetaBR_el}  \frac{ \partial [\boldsymbol{\mu}_{_\text{U},_t}]_k}{\partial{[\boldsymbol{\theta}_{_\text{BR}}]_{_i}}} &=g_{_\text{BRU}}\sqrt{(\,1-\rho)\, P_\text{B}} [\mathbf{d}(\,\tau_{_\text{BRU}} )\,]_{k}\mathbf{a}_{_\text{R}}^\top(\,\boldsymbol{\theta}_{_\text{RU}})\, \nonumber \text{diag} (\,\boldsymbol{\gamma}_{_t})\, \\&\mathbf{a}_{_\text{R}}(\,\boldsymbol{\phi}_{_\text{RB}})\,\frac{\partial{\mathbf{a}^\top_{_\text{B}}(\,\boldsymbol{\theta}_{_\text{BR}})\,}}{\partial{[\boldsymbol{\theta}_{_\text{BR}}]_{_i}}}\mathbf{f}_{_t} \quad\forall i\in \{\text{el},\text{az}\}, 
\\\label{eq:mu0_thetaBU}  \frac{ \partial [\boldsymbol{\mu}_{_\text{R},_t}]_k}{\partial{\boldsymbol{\boldsymbol{\theta}}_{_\text{BU}}}} &= \mathbf{0}^\top_{2}, \quad \text{and} \quad  \frac{ \partial [\boldsymbol{\mu}_{_\text{R},_t}]_k}{\partial{\boldsymbol{\theta}_{_\text{RU}}}} = \mathbf{0}^\top_{2},
\\\label{eq:mu0_thetaBU_el}  \frac{ \partial [\boldsymbol{\mu}_{_\text{U},_t}]_k}{\partial{[\boldsymbol{\theta}_{_\text{BU}}]_{_i}}} &= g_{_\text{BU}} \sqrt{P_\text{B}}   [\mathbf{d}(\,\tau_{_\text{BU}} )\,]_{_{k}}\frac{\partial{\mathbf{a}^\top_{_\text{B}}}(\,\boldsymbol{\theta}_{_\text{BU}})\,}{\partial{[\boldsymbol{\theta}_{_\text{BU}}]_{_i}}}\mathbf{f}_{_t} \quad\forall i\in \{\text{el},\text{az}\},
\\\label{eq:mu1_thetaRU_el}  \frac{ \partial [\boldsymbol{\mu}_{_\text{U},_t}]_k}{\partial{[\boldsymbol{\theta}_{_\text{RU}}]_{_i}}} =& g_{_\text{BRU}}\sqrt{(\,1-\rho)\, P_\text{B}} [\mathbf{d}(\,\tau_{_\text{BRU}} )\,]_{k}\frac{\partial{\mathbf{a}^\top_{_\text{R}}(\,{\boldsymbol{\theta}}_{_\text{RU}})}}{\partial{[\boldsymbol{\theta}_{_\text{RU}}]_{_i}}}\, \nonumber \text{diag} (\,\boldsymbol{\gamma}_{_t})\, \\&\mathbf{a}_{_\text{R}}(\,\boldsymbol{\phi}_{_\text{RB}})\,\mathbf{a}^\top_{_\text{B}}(\,\boldsymbol{\theta}_{_\text{BR}})\,\mathbf{f}_{_t} \quad\forall i\in \{\text{el},\text{az}\}, 
\\\label{eq:mu0_phiBR}
  \frac{ \partial [\boldsymbol{\mu}_{_\text{R},_t}]_k}{\partial{[\boldsymbol{\phi}_{_\text{RB}}]_{_i}}} =& g_{_\text{BR}} \sqrt{\rho P_\text{B}}   [\mathbf{d}(\,\tau_{_\text{BR}} )\,]_{_{k}} \mathbf{c}_{_t}^\top \frac{\partial\mathbf{a}_{_\text{R}}(\,\boldsymbol{\phi}_{_\text{RB}})\,}{\partial{[\boldsymbol{\phi}_{_\text{RB}}]_{_i}}}\mathbf{a}^\top_{_\text{B}}(\,\boldsymbol{\theta}_{_\text{BR}})\,\mathbf{f}_{_t} \nonumber \\ &\quad\forall i\in \{\text{el},\text{az}\},  
\\\label{eq:mu1_phiBR}
  \frac{ \partial [\boldsymbol{\mu}_{_\text{U},_t}]_k}{\partial{[\boldsymbol{\phi}_{_\text{RB}}}]_{_i}} =& g_{_\text{BRU}}\sqrt{(\,1-\rho)\, P_\text{B}} [\mathbf{d}(\,\tau_{_\text{BRU}} )\,]_{k}\mathbf{a}^\top_{_\text{R}}(\,\boldsymbol{\theta}_{_\text{RU}})\, \nonumber \text{diag} (\,\boldsymbol{\gamma}_{_t})\, \\&\frac{\partial\mathbf{a}_{_\text{R}}(\,\boldsymbol{\phi}_{_\text{RB}})\,}{\partial{[\boldsymbol{\phi}_{_\text{RB}}}]_{_i}}\mathbf{a}^\top_{_\text{B}}(\,\boldsymbol{\theta}_{_\text{BR}})\,\mathbf{f}_{_t} \quad\forall i\in \{\text{el},\text{az}\}, 
\\\label{eq:mu0_gBR_real}
  \frac{ \partial [\boldsymbol{\mu}_{_\text{R},_t}]_k}{\partial{\mathbf{g}_{_\text{BR}}}} =& \mathbf{e}^\top \otimes \sqrt{\rho P_\text{B}}   [\mathbf{d}(\,\tau_{_\text{BR}} )\,]_{_{k}} \mathbf{c}_{_t}^\top\mathbf{a}_{_\text{R}}(\,\boldsymbol{\phi}_{_\text{RB}})\,\mathbf{a}^\top_{_\text{B}}(\,\boldsymbol{\theta}_{_\text{BR}})\,\mathbf{f}_{_t} ,  
\\\label{eq:mu0_gBU}
  \frac{ \partial [\boldsymbol{\mu}_{_\text{U},_t}]_k}{\partial{\mathbf{g}_{_\text{BR}}}} &= \mathbf{0}_{_2}^\top, \frac{ \partial [\boldsymbol{\mu}_{_\text{R},_t}]_k}{\partial{\mathbf{g}_{_\text{BU}}}} = \mathbf{0}_{_2}^\top,   \frac{ \partial [\boldsymbol{\mu}_{_\text{R},_t}]_k}{\partial{\mathbf{g}_{_\text{BU}}}} = \mathbf{0}_{_2}^\top,
\\\label{eq:mu1_gBU_real}
  \frac{ \partial [\boldsymbol{\mu}_{_\text{U},_t}]_k}{\partial{\mathbf{g}_{_\text{BU}}}} &= \mathbf{e}^\top \otimes \sqrt{\rho P_\text{B}}   [\mathbf{d}(\,\tau_{_\text{BU}} )\,]_{_{k}} \mathbf{a}^\top_{_\text{B}}(\,\boldsymbol{\theta}_{_\text{BU}})\,\mathbf{f}_{_t},
\\\label{eq:mu1_gBRU_imag}
  \frac{ \partial [\boldsymbol{\mu}_{_\text{U},_t}]_k}{\partial  \mathbf{g}_{_\text{BRU}}} =& \mathbf{e}^\top \otimes \sqrt{(\,1-\rho)\, P_\text{B}} [\mathbf{d}(\,\tau_{_\text{BRU}} )\,]_{k}\mathbf{a}^\top_{_\text{R}}(\,\boldsymbol{\theta}_{_\text{RU}})\, \nonumber \text{diag} (\,\boldsymbol{\gamma}_{_t})\, \\&\mathbf{a}_{_\text{R}}(\,\boldsymbol{\phi}_{_\text{RB}})\,\mathbf{a}_{_\text{B}}^\top(\,\boldsymbol{\theta}_{_\text{BR}})\,\mathbf{f}_{_t} , 
\end{align}
\end{subequations}
where $\mathbf{\dot{d}}(\,\tau )\, \triangleq -\jmath 2\pi\Delta_f [0,\dots, k-1]^\top \odot \mathbf{d}(\,\tau )\,$, $\mathbf{e} \triangleq [1,\jmath]^\top$, and $\forall i\in \{\text{el},\text{az}\}$:
\begin{align}
\allowdisplaybreaks
\label{eq:driv_a_psi}
\frac{\partial\mathbf{a}(\,\boldsymbol{\psi})\,}{\partial[\boldsymbol{\psi}]_{_i}}&= \frac{\partial\mathbf{a}_r(\boldsymbol{\psi})}{\partial[\boldsymbol{\psi}]_{_i}}\otimes \mathbf{a}_c(\boldsymbol{\psi})+\mathbf{a}_r(\boldsymbol{\psi})\otimes\frac{\partial\mathbf{a}_c(\boldsymbol{\psi})}{\partial[\boldsymbol{\psi}]_{_i}}.
\end{align} 
By using~\eqref{eq:ar} and \eqref{eq:ac}, and the notation $\mathbf{m}\triangleq[-(M-1)/2, \dots,(M-1)/2]^\top$ with $M\in \{M_{_\text{B}},M_{_\text{R}}\}$, we can write: 
\begin{subequations}
\allowdisplaybreaks
\begin{align}
\frac{\partial\mathbf{a}_r(\boldsymbol{\psi})}{\partial[\boldsymbol{\psi}]_{_\text{el}}}&= -\jmath \frac{2\pi d}{\lambda} \cos{[\boldsymbol{\psi}]_\text{el}}\cos{[\boldsymbol{\psi}]_\text{az}}\mathbf{m}\odot\mathbf{a}_r(\boldsymbol{\psi}), \\
\frac{\partial\mathbf{a}_r(\boldsymbol{\psi})}{\partial[\boldsymbol{\psi}]_{_\text{az}}}&= \jmath \frac{2\pi d}{\lambda} \sin{[\boldsymbol{\psi}]_\text{el}}\sin{[\boldsymbol{\psi}]_\text{az}}\mathbf{m}\odot\mathbf{a}_r(\boldsymbol{\psi}), \\
\frac{\partial\mathbf{a}_c(\boldsymbol{\psi})}{\partial[\boldsymbol{\psi}]_{_\text{el}}}&= \jmath \frac{2\pi d}{\lambda} \sin{[\boldsymbol{\psi}]_\text{el}}\mathbf{m}\odot\mathbf{a}_c(\boldsymbol{\psi}),\\
\frac{\partial\mathbf{a}_c(\boldsymbol{\psi})}{\partial[\boldsymbol{\psi}]_{_\text{az}}}&= \mathbf{0}_{_M}.
\end{align}
\end{subequations}

\section{Derivation of $\mathbf{T}$}\label{Sec.T}
Using \eqref{eq:tau} and ~\eqref{eq:tau_sp0}, the elements of $\mathbf{T}$ are computed as:
\begin{subequations}\label{eq:dtau}
\begin{align}\label{eq:dtau-BR}
  \mathbf{T}(1,1:3) &= \frac{ \partial\tau_{_\text{BR}}} {\partial \mathbf{p}_\text{R}} = \frac{\mathbf{p}_{\text{R}}-\mathbf{p}_{\text{B}}}{c\Vert\mathbf{p}_{\text{B}}-\mathbf{p}_{\text{R}}\Vert} , 
\\\label{eq:dtau-clkR}
  \mathbf{T}(1,7) &= \frac{ \partial\tau_{_\text{BR}}} {\partial b_{\text{R}}} = 1, \quad \text{and} \quad \mathbf{T}(2,8) = \frac{ \partial\tau_{_\text{BU}}} {\partial b_{\text{U}}} = 1, 
\\\label{eq:dtau-BR}
  \mathbf{T}(2,4:6) &= \frac{ \partial\tau_{_\text{BU}}} {\partial \mathbf{p}_\text{U}} = \frac{\mathbf{p}_{\text{U}}-\mathbf{p}_{\text{B}}}{c\Vert\mathbf{p}_{\text{B}}-\mathbf{p}_{\text{U}}\Vert} , 
\\
\label{eq:dtaubru-BR}
  \mathbf{T}(3,1:3) &= \frac{ \partial\tau_{_\text{BRU}}} {\partial \mathbf{p}_\text{R}} = \frac{\mathbf{p}_{\text{R}}-\mathbf{p}_{\text{B}}}{c\Vert\mathbf{p}_{\text{B}}-\mathbf{p}_{\text{R}}\Vert} , 
\\\label{eq:dtaubru-BR}
  \mathbf{T}(3,4:6) &= \frac{ \partial\tau_{_\text{BRU}}} {\partial \mathbf{p}_\text{U}} = \frac{\mathbf{p}_{\text{U}}-\mathbf{p}_{\text{B}}}{c\Vert\mathbf{p}_{\text{B}}-\mathbf{p}_{\text{U}}\Vert} , 
\\\label{eq:dtau-clkuR}
  \mathbf{T}(3,7:8) &= \Big[\,\frac{ \partial\tau_{_\text{BRU}}} {\partial b_{\text{R}}},\frac{ \partial\tau_{_\text{BRU}}} {\partial b_{\text{U}}}\Big]\, = [1,1]. 
\end{align}
\end{subequations}
To derive the derivatives of the AOAs and AODs w.r.t. state parameters, we first introduce the following auxiliary variables:
\begin{subequations}
\begin{equation}\label{eq:ud_B}
 \mathbf{u}_{_\text{DR}}\triangleq \frac{\mathbf{p}_\text{R}-\mathbf{p}_\text{B}}{\Vert\mathbf{p}_\text{R}-\mathbf{p}_\text{B}\Vert}\quad \text{and } \quad   \mathbf{u}_{_\text{DU}}\triangleq \frac{\mathbf{p}_\text{U}-\mathbf{p}_\text{B}}{\Vert\mathbf{p}_\text{U}-\mathbf{p}_\text{B}\Vert},  
\end{equation}
\begin{equation}\label{eq:u_R}
 \mathbf{v}_{_\text{AB}}\triangleq \frac{\mathbf{p}_\text{B}-\mathbf{p}_\text{R}}{\Vert\mathbf{p}_\text{B}-\mathbf{p}_\text{R}\Vert}\quad \text{and } \quad   \mathbf{v}_{_\text{DU}}\triangleq \frac{\mathbf{p}_\text{U}-\mathbf{p}_\text{R}}{\Vert\mathbf{p}_\text{U}-\mathbf{p}_\text{R}\Vert},  
\end{equation}
\end{subequations}
as well as $\mathbf{u}_{_1}\triangleq[1,0,0]^\top$, $\mathbf{u}_{_2}\triangleq[0,1,0]^\top$, and $\mathbf{u}_{_3}\triangleq[0,0,1]^\top$. 
Then, we may rewrite the AOAs and AODs as follows~\cite[Appendix A]{nazari2023mmwave}:
\begin{subequations}\label{eq:AODs-BS_re}
\allowdisplaybreaks
\begin{align}\label{eq:AODs-BS-RIS_re}
\boldsymbol{\theta}_{_\text{BR}}&=[     {\theta}_{_\text{BR}}^{(\text{az})},
{\theta}_{_\text{BR}}^{(\text{el})}
]^\top=[\text{atan2}(\mathbf{u}_2^\top\mathbf{u}_{_\text{DR}},\mathbf{u}_1^\top\mathbf{u}_{_\text{DR}}),
\text{acos}(\mathbf{u}_3^\top\mathbf{u}_{_\text{DR}})]^\top,\\ \label{eq:AODs-BS-UE_re}
\boldsymbol{\theta}_{_\text{BU}}&=[
{\theta}_{_\text{BU}}^{(\text{az})},
{\theta}_{_\text{BU}}^{(\text{el})}
]^\top=[\text{atan2}(\mathbf{u}_2^\top\mathbf{u}_{_\text{DU}},\mathbf{u}_1^\top\mathbf{u}_{_\text{DU}}),
\text{acos}(\mathbf{u}_3^\top\mathbf{u}_{_\text{DR}})]^\top,\\\label{eq:AOAs-RIS-BS_re}
\boldsymbol{\phi}_{_\text{RB}}&= [
{\phi}_{_\text{RB}}^{(\text{az})},
{\phi}_{_\text{RB}}^{(\text{el})}
]^\top=[\text{atan2}(\mathbf{r}_2^\top\mathbf{v}_{_\text{AB}}),\mathbf{r}_1^\top\mathbf{v}_{_\text{AB}})),
\text{acos}(\mathbf{r}_2^\top\mathbf{v}_{_\text{AB}})]^\top,\\ \label{eq:AODs-RIS-UE_re}
\boldsymbol{\theta}_{_\text{RU}}&=     [{\theta}_{_\text{RU}}^{(\text{az})},
{\theta}_{_\text{RU}}^{(\text{el})}]^\top= [
\text{atan2}(\mathbf{r}_2^\top\mathbf{v}_{_\text{DU}},\mathbf{r}_1^\top\mathbf{v}_{_\text{DU}}),
\text{acos}(\mathbf{r}_3^\top\mathbf{v}_{_\text{DU}})]^\top, 
\end{align}
\end{subequations}
yielding the following derivatives:
\begin{subequations}
\allowdisplaybreaks
\begin{align}\label{eq:partial_uB_pris}
  \frac{\partial \mathbf{u}_{_\text{DR}}}{\partial\mathbf{p}_\text{R} }&= (\,\mathbf{I}_3- \mathbf{u}_{_\text{DR}}  \mathbf{u}_{_\text{DR}}^\top)\,/\Vert\mathbf{p}_\text{B}-\mathbf{p}_\text{R}\Vert\quad,
\\\label{eq:partial_uB_pu}
  \frac{\partial \mathbf{u}_{_\text{DU}}}{\partial\mathbf{p}_\text{U} } &= (\,\mathbf{I}_3- \mathbf{u}_{_\text{DU}}  \mathbf{u}_{_\text{DU}}^\top)\,/\Vert\mathbf{p}_\text{B}-\mathbf{p}_\text{U}\Vert\quad,
\\\label{eq:partial_v_pris}
  \frac{\partial \mathbf{v}_{_\text{AB}}}{\partial\mathbf{p}_\text{R} }& = (\, \mathbf{v}_{_\text{AB}}  \mathbf{v}_{_\text{AB}}^\top-\mathbf{I}_3)\,/\Vert\mathbf{p}_\text{B}-\mathbf{p}_\text{R}\Vert\quad,
\\\label{eq:partial_vdu_pris}
  \frac{\partial \mathbf{v}_{_\text{DU}}}{\partial\mathbf{p}_\text{R} } &= (\, \mathbf{v}_{_\text{DU}}  \mathbf{v}_{_\text{DU}}^\top-\mathbf{I}_3)\,/\Vert\mathbf{p}_\text{U}-\mathbf{p}_\text{R}\Vert\quad,
\\\label{eq:partial_v_pu}
  \frac{\partial \mathbf{v}_{_\text{DU}}}{\partial\mathbf{p}_\text{U} } &= (\,\mathbf{I}_3- \mathbf{v}_{_\text{DU}}  \mathbf{v}_{_\text{DU}}^\top)\,/\Vert\mathbf{p}_\text{U}-\mathbf{p}_\text{R}\Vert\quad,
\\\label{eq:partial_tethaBR_az_UD}
    \frac{\partial {\theta}_{_\text{BR}}^{(\text{az})}}{\partial\mathbf{u}_{_\text{DR}}} &=\frac{(\,\mathbf{u}_1^\top\mathbf{u}_{_\text{DR}})\,\mathbf{u}_2 - (\,\mathbf{u}_2^\top\mathbf{u}_{_\text{DR}})\,\mathbf{u}_1}{(\,\mathbf{u}_1^\top\mathbf{u}_{_\text{DR}})\,^2 + (\,\mathbf{u}_2^\top\mathbf{u}_{_\text{DR}})\,^2}\quad,
\\\label{eq:partial_tethaBR_el_UD}
    \frac{\partial {\theta}_{_\text{BR}}^{(\text{el})}}{\partial\mathbf{u}_{_\text{DR}}} &=-\frac{\mathbf{u}_3}{\sqrt{1-(\mathbf{u}_3^\top\mathbf{u}_{_\text{DR}})^2}}\quad,
\\\label{eq:partial_tethaBU_az_UD}
    \frac{\partial {\theta}_{_\text{BU}}^{(\text{az})}}{\partial\mathbf{u}_{_\text{DR}}} &=\frac{(\,\mathbf{u}_1^\top\mathbf{u}_{_\text{DR}})\,\mathbf{u}_2 - (\,\mathbf{u}_2^\top\mathbf{u}_{_\text{DR}})\,\mathbf{u}_1}{(\,\mathbf{u}_1^\top\mathbf{u}_{_\text{DR}})\,^2 + (\,\mathbf{u}_2^\top\mathbf{u}_{_\text{DR}})\,^2}\quad,
\\\label{eq:partial_tethaBU_a_UD}
    \frac{\partial {\theta}_{_\text{BU}}^{(\text{el})}}{\partial\mathbf{u}_{_\text{DR}}} &=-\frac{\mathbf{u}_3}{\sqrt{1-(\mathbf{u}_3^\top\mathbf{u}_{_\text{DR}})^2}}\quad,
\\\label{eq:partial_thetaRU_uDR}
     \frac{\partial {\theta}_{_\text{RU}}^{(\text{az})}}{\partial\mathbf{v}_{_\text{DU}}}&= \frac{(\,\mathbf{r}_1^\top\mathbf{v}_{_\text{DU}})\,\mathbf{r}_2 - (\,\mathbf{r}_2^\top\mathbf{v}_{_\text{DU}})\,\mathbf{r}_1}{(\,\mathbf{r}_1^\top\mathbf{v}_{_\text{DU}})\,^2 + (\,\mathbf{r}_2^\top\mathbf{v}_{_\text{DU}})\,^2}\quad,
\\\label{eq:partial_tethaRU_el_UD}
    \frac{\partial {\theta}_{_\text{RU}}^{(\text{el})}}{\partial\mathbf{v}_{_\text{DU}}} &=-\frac{\mathbf{r}_3}{\sqrt{1-(\mathbf{r}_3^\top\mathbf{v}_{_\text{DU}})^2}}\quad,
\\
\label{eq:partial_phiBR_az_uAB}
     \frac{\partial {\phi}_{_\text{RB}}^{(\text{az})}}{\partial\mathbf{v}_{_\text{AB}}}&= \frac{(\,\mathbf{r}_1^\top\mathbf{v}_{_\text{AB}})\,\mathbf{r}_2 - (\,\mathbf{r}_2^\top\mathbf{v}_{_\text{AB}})\,\mathbf{r}_1}{(\,\mathbf{r}_1^\top\mathbf{v}_{_\text{AB}})\,^2 + (\,\mathbf{r}_2^\top\mathbf{v}_{_\text{AB}})\,^2}\quad,\\
\label{eq:partial_phiRB_az_UD}
    \frac{\partial {\phi}_{_\text{RB}}^{(\text{el})}}{\partial\mathbf{v}_{_\text{AB}}} &=-\frac{\mathbf{r}_3}{\sqrt{1-(\mathbf{r}_3^\top\mathbf{v}_{_\text{AB}})^2}}\quad,\\
\label{eq:partial_phiRB_az_R}
    \frac{\partial {\phi}_{_\text{RB}}^{(\text{az})}}{\partial\mathbf{R}}&= \frac{(\,\mathbf{r}_1^\top\mathbf{v}_{_\text{AB}})\,\mathbf{v}_{_\text{AB}}\mathbf{u}_2^\top - (\,\mathbf{r}_2^\top\mathbf{v}_{_\text{AB}})\,\mathbf{v}_{_\text{AB}}\mathbf{u}_1^\top}{(\,\mathbf{r}_1^\top\mathbf{v}_{_\text{AB}})\,^2 + (\,\mathbf{r}_2^\top\mathbf{v}_{_\text{AB}})\,^2}\quad,
    \\
\label{eq:partial_phiRB_el_R}
    \frac{\partial {\phi}_{_\text{RB}}^{(\text{el})}}{\partial\mathbf{R}} &=-\frac{\mathbf{v}_{_\text{AB}}\mathbf{u}_3^\top}{\sqrt{1-(\mathbf{r}_3^\top\mathbf{v}_{_\text{AB}})^2}}\quad,\\
\label{eq:partial_thetaRU_az_R}
    \frac{\partial {\theta}_{_\text{RU}}^{(\text{az})}}{\partial\mathbf{R}}&= \frac{(\,\mathbf{r}_1^\top\mathbf{v}_{_\text{DU}})\,\mathbf{v}_{_\text{DU}}\mathbf{u}_2^\top - (\,\mathbf{r}_2^\top\mathbf{v}_{_\text{DU}})\,\mathbf{v}_{_\text{DU}}\mathbf{u}_1^\top}{(\,\mathbf{r}_1^\top\mathbf{v}_{_\text{DU}})\,^2 + (\,\mathbf{r}_2^\top\mathbf{v}_{_\text{DU}})\,^2}\quad,
    \\
\label{eq:partial_thetaRU_el_R}
    \frac{\partial {\theta}_{_\text{RU}}^{(\text{el})}}{\partial\mathbf{R}} &=-\frac{\mathbf{v}_{_\text{DU}}\mathbf{u}_3^\top}{\sqrt{1-(\mathbf{r}_3^\top\mathbf{v}_{_\text{DU}})^2}}\quad,
\end{align}
\end{subequations}
The latter expressions are used to compute the following elements of $\mathbf{T}$ (with those remaining being zero):
\begin{subequations}
\allowdisplaybreaks
\begin{align}\label{eq:dtheta_BR_rris}
  \mathbf{T}(4,1:3) &= \frac{ \partial\theta_{_\text{BR}}^{(\text{az})}} {\partial \mathbf{p}_\text{R}} = \frac{\partial\theta_{_\text{BR}}^{(\text{az})}}{\partial\mathbf{u}_{_\text{DR}}} \frac{\partial \mathbf{u}_{_\text{DR}}}{\partial\mathbf{p}_\text{R} } , 
\\\label{eq:dtheta_BR_el_rris}
  \mathbf{T}(5,1:3) &= \frac{ \partial\theta_{_\text{BR}}^{(\text{el})}} {\partial \mathbf{p}_\text{R}} = \frac{\partial\theta_{_\text{BR}}^{(\text{el})}}{\partial\mathbf{u}_{_\text{DR}}} \frac{\partial \mathbf{u}_{_\text{DR}}}{\partial\mathbf{p}_\text{R} } , 
\\\label{eq:dtheta_BU_az_Pu}
  \mathbf{T}(6,4:6) &= \frac{ \partial\theta_{_\text{BU}}^{(\text{az})}} {\partial \mathbf{p}_\text{U}} = \frac{\partial\theta_{_\text{BU}}^{(\text{az})}}{\partial\mathbf{u}_{_\text{DU}}} \frac{\partial \mathbf{u}_{_\text{DU}}}{\partial\mathbf{p}_\text{U} } , 
\\\label{eq:dtheta_BU_el_Pu}
  \mathbf{T}(7,4:6) &= \frac{ \partial\theta_{_\text{BU}}^{(\text{el})}} {\partial \mathbf{p}_\text{U}} = \frac{\partial\theta_{_\text{BU}}^{(\text{el})}}{\partial\mathbf{u}_{_\text{DU}}} \frac{\partial \mathbf{u}_{_\text{DU}}}{\partial\mathbf{p}_\text{U} } , 
\\\label{eq:dtheta_RU_az_Pris}
  \mathbf{T}(8,1:3) &= \frac{ \partial\theta_{_\text{RU}}^{(\text{az})}} {\partial \mathbf{p}_\text{R}} = \frac{\partial\theta_{_\text{RU}}^{(\text{az})}}{\partial\mathbf{v}_{_\text{DU}}} \frac{\partial \mathbf{v}_{_\text{DU}}}{\partial\mathbf{p}_\text{R} } , 
\\\label{eq:dtheta_RU_el_Pris}
  \mathbf{T}(9,1:3) &= \frac{ \partial\theta_{_\text{RU}}^{(\text{el})}} {\partial \mathbf{p}_\text{R}} = \frac{\partial\theta_{_\text{RU}}^{(\text{el})}}{\partial\mathbf{v}_{_\text{DU}}} \frac{\partial \mathbf{v}_{_\text{DU}}}{\partial\mathbf{p}_\text{R} } , 
\\\label{eq:dtheta_RU_az_Pu}
  \mathbf{T}(8,4:6) &= \frac{ \partial\theta_{_\text{RU}}^{(\text{az})}} {\partial \mathbf{p}_\text{U}} = \frac{\partial\theta_{_\text{RU}}^{(\text{az})}}{\partial\mathbf{v}_{_\text{DU}}} \frac{\partial \mathbf{v}_{_\text{DU}}}{\partial\mathbf{p}_\text{U} } , 
\\\label{eq:dtheta_RU_el_Pu}
  \mathbf{T}(9,4:6) &= \frac{ \partial\theta_{_\text{RU}}^{(\text{el})}} {\partial \mathbf{p}_\text{U}} = \frac{\partial\theta_{_\text{RU}}^{(\text{el})}}{\partial\mathbf{v}_{_\text{DU}}} \frac{\partial \mathbf{v}_{_\text{DU}}}{\partial\mathbf{p}_\text{U} } , 
\\\label{eq:dphiBR_az_PR}
  \mathbf{T}(10,1:3) &= \frac{ \partial\phi_{_\text{RB}}^{(\text{az})}} {\partial \mathbf{p}_\text{R}} = \frac{\partial\phi_{_\text{RB}}^{(\text{az})}}{\partial\mathbf{v}_{_\text{AB}}} \frac{\partial \mathbf{v}_{_\text{AB}}}{\partial\mathbf{p}_\text{R} } , 
\\\label{eq:dphiBR_el_PR}
  \mathbf{T}(11,1:3) &= \frac{ \partial\phi_{_\text{RB}}^{(\text{el})}} {\partial \mathbf{p}_\text{R}} = \frac{\partial\phi_{_\text{RB}}^{(\text{el})}}{\partial\mathbf{v}_{_\text{AB}}} \frac{\partial \mathbf{v}_{_\text{AB}}}{\partial\mathbf{p}_\text{R} } , 
\\\label{eq:partial_thetaRU_R}
    \mathbf{T}(8,9:17)& =\frac{\partial {\theta}_{_\text{RU}}^{(\text{az})}}{\partial\mathbf{R}}, \quad \text{and} \quad  \mathbf{T}(9,9:17) =\frac{\partial {\theta}_{_\text{RU}}^{(\text{el})}}{\partial\mathbf{R}},
\\\label{eq:partial_phiRB_R}
    \mathbf{T}(10,9:17) &=\frac{\partial {\phi}_{_\text{RB}}^{(\text{az})}}{\partial\mathbf{R}}, \quad \text{and} \quad  \mathbf{T}(11,9:17) =\frac{\partial {\phi}_{_\text{RB}}^{(\text{el})}}{\partial\mathbf{R}}.
\end{align}
\end{subequations}
\balance 
\bibliographystyle{IEEEtran}
\bibliography{ref_revise.bib}
\end{document}